\documentclass[dvips]{acta}
\usepackage{supertabular,lscape,epsfig}
\usepackage{amssymb}
\usepackage{amsmath}
\usepackage{multirow}
\usepackage{placeins}
\DeclareSymbolFont{ppa}{OT1}{ppl}{m}{it}
\DeclareMathSymbol{\vv}{\mathalpha}{ppa}{'166}

\newfont{\hb}{rphvb at 10pt}
\newfont{\hbo}{rphvbo at 10pt}
\newfont{\bitt}{rptmbi at 12pt}
\newfont{\bits}{rptmbi at 11pt}

\SetPages{269}{292}

\SetVol{66}{2016}

\begin{document}

\newcommand{\TabApp}[2]{\begin{center}\parbox[t]{#1}{\centerline{
  {\bf Appendix}}
  \vskip2mm
  \centerline{\small {\spaceskip 2pt plus 1pt minus 1pt T a b l e}
  \refstepcounter{table}\thetable}
  \vskip2mm
  \centerline{\footnotesize #2}}
  \vskip3mm
\end{center}}

\newcommand{\TabCapp}[2]{\begin{center}\parbox[t]{#1}{\centerline{
  \small {\spaceskip 2pt plus 1pt minus 1pt T a b l e}
  \refstepcounter{table}\thetable}
  \vskip2mm
  \centerline{\footnotesize #2}}
  \vskip3mm
\end{center}}

\newcommand{\TTabCap}[3]{\begin{center}\parbox[t]{#1}{\centerline{
  \small {\spaceskip 2pt plus 1pt minus 1pt T a b l e}
  \refstepcounter{table}\thetable}
  \vskip2mm
  \centerline{\footnotesize #2}
  \centerline{\footnotesize #3}}
  \vskip1mm
\end{center}}

\newcommand{\MakeTableApp}[4]{\begin{table}[p]\TabApp{#2}{#3}
  \begin{center} \TableFont \begin{tabular}{#1} #4 
  \end{tabular}\end{center}\end{table}}

\newcommand{\MakeTableSepp}[4]{\begin{table}[p]\TabCapp{#2}{#3}
  \begin{center} \TableFont \begin{tabular}{#1} #4 
  \end{tabular}\end{center}\end{table}}

\newcommand{\MakeTableee}[4]{\begin{table}[htb]\TabCapp{#2}{#3}
  \begin{center} \TableFont \begin{tabular}{#1} #4
  \end{tabular}\end{center}\end{table}}

\newcommand{\MakeTablee}[5]{\begin{table}[htb]\TTabCap{#2}{#3}{#4}
  \begin{center} \TableFont \begin{tabular}{#1} #5 
  \end{tabular}\end{center}\end{table}}

\newfont{\bb}{ptmbi8t at 12pt}
\newfont{\bbb}{cmbxti10}
\newfont{\bbbb}{cmbxti10 at 9pt}
\newcommand{\uprule}{\rule{0pt}{2.5ex}}
\newcommand{\douprule}{\rule[-2ex]{0pt}{4.5ex}}
\newcommand{\dorule}{\rule[-2ex]{0pt}{2ex}}
\begin{Titlepage}
\Title{OGLE-ing the Magellanic System:\\ Photometric Metallicity from Fundamental Mode RR~Lyrae Stars}
\Author{D.\,M.~~S~k~o~w~r~o~n$^1$,~~
I.~~S~o~s~z~y~ñ~s~k~i$^1$,~~
A.~~U~d~a~l~s~k~i$^1$,~~
M.\,K.~~S~z~y~m~a~ñ~s~k~i$^1$,\\
P.~~P~i~e~t~r~u~k~o~w~i~c~z$^1$,~~
J.~~S~k~o~w~r~o~n$^1$,~~
R.~~P~o~l~e~s~k~i$^{1,2}$,~~
£.~~W~y~r~z~y~k~o~w~s~k~i$^1$,\\
K.~~U~l~a~c~z~y~k$^{1,3}$,~~
S.~~K~o~z~³~o~w~s~k~i$^1$,~~
P.~~M~r~ó~z$^1$~~ and~~
M.~~P~a~w~l~a~k$^1$}
{$^1$ Warsaw University Observatory, Al.~Ujazdowskie~4, 00-478~Warszawa,
Poland\\
$^2$ Department of Astronomy, Ohio State University, 140 W. 18th Ave.,
Columbus, OH~43210,~USA\\
$^3$ Department of Physics, University of Warwick, Gibbet Hill Road,
Coventry, CV4~7AL,~UK\\
e-mail: dszczyg@astrouw.edu.pl}

\Received{August 1, 2016}
\end{Titlepage}

\Abstract{In an era of extensive photometric observations, the catalogs of
  RR~Lyr type variable stars number tens of thousands of objects. The
  relation between the iron abundance [Fe/H] and the Fourier parameters of
  the stars light curve allows us to investigate mean metallicities and
  metallicity gradients in various stellar environments, independently of
  time-consuming spectroscopic observations. In this paper we use almost
  6500 {\it V}- and {\it I}-band light curves of fundamental mode RR~Lyr
  stars from the OGLE-IV survey to provide a relation between the {\it V}-
  and {\it I}-band phase parameter $\varphi_{31}$ used to estimate [Fe/H].
  The relation depends on metallicity, which limits its applicability. We
  apply this relation to metallicity formulae developed for the Johnson
  {\it V}- and the Kepler {\it Kp}-band to obtain the relation between
  [Fe/H] and $\varphi_{31}$ for the {\it I}-band photometry. Last, we apply
  the new relation of Nemec to the OGLE-IV fundamental mode RR~Lyr stars
  data and construct a metallicity map of the Magellanic Clouds. Median
  [Fe/H] is $-1.39\pm0.44$~dex for the LMC and $-1.77\pm0.48$~dex for the
  SMC, on the Jurcsik metallicity scale. We also find a metallicity
  gradient within the LMC with a slope of $-0.029\pm0.002$~dex/kpc in the
  inner 5~kpc and $-0.030 \pm0.003$~dex/kpc beyond 8~kpc, and no gradient
  in-between ($-0.019\pm0.002$~dex/kpc integrally). We do not observe a
  metallicity gradient in the SMC, although we show that the metal-rich
  RRab stars are more concentrated toward the SMC center than the
  metal-poor.}{Stars: abundances -- Stars: statistics -- Stars: variables:
  RR~Lyr -- Galaxies: abundances -- Magellanic Clouds}

\Section{Introduction}
The importance of RR~Lyr (RRL) type variable stars in modern astrophysics
cannot be underestimated: they serve as standard candles in our Galaxy and
in the Local Group, they trace the Galactic structure and evolution, they
represent the old and metal-poor population that gives insight into the
early chemical composition of the nearby Universe.

Physical parameters of the RRL pulsating variables are reflected in the
characteristic shape of their light curves, that can be well modeled with
the Fourier series. Several studies have shown that there exists a relation
between the Fourier parameters and the iron abundance [Fe/H] of RRL stars
pulsating in the fundamental mode (RRab), and we will describe them in
Section~2. Such a relation is of great importance in an era where
photometric databases are orders of magnitude larger than spectroscopic
ones, and we count RRL light curves in tens of thousands (Soszyñski \etal
2014, 2016). This allows for studying the metal content of galaxies
independently of spectroscopic observations which are much more
time-consuming and difficult to collect, and in some cases even impossible
to obtain. Large photometric databases of RRL stars have already been used
to investigate the chemical characteristics of the Galactic bulge
(Pietrukowicz \etal 2012, 2015, Gonzalez \etal 2013, Sans Fuentes and De
Ridder 2014) and field (Kinemuchi \etal 2006, Szczygie³ \etal 2009,
Torrealba \etal 2015), and the Magellanic Clouds (Deb and Singh 2010, 2014,
Deb \etal 2015, Feast \etal 2010, Haschke \etal 2012, Kapakos \etal
2011, 2012; Wagner-Kaiser and Sarajedini 2013).

In this paper, we take a look at available methods of determining [Fe/H]
from the light curve parameters, and by determining a transformation
between the {\it I}- and the {\it V}-band phase parameters, we present a
relation for an accurate [Fe/H] calculation from the {\it I}-band RRab
data. For this we use a sample of almost 6500 RRab stars from the fourth
phase of the Optical Gravitational Lensing Experiment (OGLE-IV, Udalski,
Szymañski and Szymañski 2015) with well observed light curves.  We then
apply the relation to the newest collection of RR~Lyr type variables in the
Magellanic Clouds (Soszyñski \etal 2016) and construct a metallicity map of
the Magellanic System.

It is worth noting, that the empirical relations presented in Section~2
should not be used for estimating [Fe/H] for individual stars, but rather
for statistically large datasets. For example, they can be used for
determining mean iron abundances of globular clusters, metallicity
gradients within galaxies, or for comparing metal content of different
stellar structures.

\Section{Photometric Metallicity Calculation Methods for RRab stars}
The first relation between the star's iron abundance and its light curve
parameters was introduced by Kov\'acs and Zsoldos (1995) and later improved
by Jurcsik and Kov\'acs (1996, hereafter JK96). In this method the
metallicity is calculated from the stellar pulsation period $P$ and the
phase combination $\varphi_{31}=\varphi_{3}-3\varphi_{1}$ from a sine
series Fourier fit:
$${\rm [Fe/H]}_{\it JK}=-5.038-5.394P+1.345\varphi_{31}.\eqno(1)$$

The formula was calibrated with 81 Galactic RRab stars that had accurate
spectroscopic metallicities and well covered Johnson {\it V}-band light
curves, and then tested on six Galactic and five Large Magellanic Cloud
(LMC) globular clusters. The overall agreement of photometric and
spectroscopic [Fe/H] values was good, but there is a systematic difference
at the low metallicity end (${\rm [Fe/H]<-2.0}$~dex) such that the
photometric metallicities have higher values (by $\approx0.3~$dex).

Another method for calculating iron abundances from RRab light curves was
proposed by Sandage (2004), where [Fe/H] is calculated from the star's
period and {\it V}-band light curve amplitude:
$${\rm [Fe/H]}_{\it Sandage}=-1.453A_V-7.990\log P-2.145.\eqno(2)$$

This formula is less sensitive to light curve quality, but as shown by
Szczygie³ \etal (2009), using 1227 All Sky Automated Survey (ASAS) RRab
light curves, it has much larger scatter when compared to the JK96
method. The authors also show that it produces an unphysical bimodal
distribution of photometric metallicities, which is a reflection of the
Oosterhoff dichotomy, and is not seen with spectroscopically obtained
[Fe/H] values. Alcock \etal (2000) provide a similar relation, that uses
the star's period and amplitude, thus it is also subject to larger [Fe/H]
scatter and the dependence on the Oosterhoff dichotomy.

The release of large catalogs of fine quality RRL light curves in the
Cousins {\it I}-band by the OGLE group, encouraged Smolec (2005, hereafter
S05) to estimate a new calibration of the $(P, \varphi_{31})\rightarrow
{\rm [Fe/H]}$ relation specifically for the {\it I}-band. It was based on
28 field RRab stars for which both spectroscopic and photometric data were
available and is valid for Fourier sine series fit:
\setcounter{equation}{2}
\begin{equation}
{\setlength\arraycolsep{2pt}
\begin{array}{ccr@{.}lcr@{.}lcr@{.}lcc}
{\rm [Fe/H]} = &-  & 3&142 &-  & 4&902P& + &0&824\phi_{31} &\\ 
               &\pm& 0&646 &\pm& 0&375 &\pm&0&104  &\quad \sigma=0.18
\end{array}}
\end{equation}

This formula was calibrated in the metallicity range $-1.7<{\rm [Fe/H]}<0.3$~dex.

Another attempt to determine [Fe/H] from the {\it I}-band photometry was made
by Deb and Singh (2010). The authors used good quality {\it V}-and {\it I}-band
light curves of 29 stars in M3, to derive $V\leftrightarrow I$ interrelations
between Fourier parameters $\varphi_{21}$, $\varphi_{31}$, and $\varphi_{41}$
(from a cosine series fit), in order to be able to determine physical parameters
of 352 RRL from OGLE-II {\it I}-band data for the Small Magellanic Cloud (SMC),
which was much higher quality that the {\it V}-band data. In particular,
they found a linear relation between $\varphi^V_{31}$ and $\varphi^I_{31}$
for RRab stars in the form:
\begin{equation}
{\setlength\arraycolsep{2pt}
\begin{array}{ccr@{.}lcc}
\varphi^V_{31}= &    &0&568 \varphi^I_{31} &+  &0.436 \\
                &\pm &0&030                &\pm&0.075 \\
\end{array}}
\end{equation}
and later used it together with the JK96 method to calculate photometric
metallicities.

Deb and Singh (2014) compared metallicity values ${\rm [Fe/H]}_S$ derived
from the {\it I}-band light curves by S05 (Eq.~3), with metallicity values
${\rm [Fe/H]}_{\it JK}$ (Eq.~1), after converting $\varphi^I_{31}$ to
$\varphi^V_{31}$ (Eq.~4). The comparison was done on a sample of 13\,095
OGLE-III RRab light curves from the LMC and revealed that ${\rm [Fe/H]}_S$
values are on average 0.07~dex higher than ${\rm [Fe/H]}_{\it JK}$ values. The
authors do not comment on this result, but possible explanations may
include the small number of stars (29) used to derive Eq.(4), the small
number of stars (28) used to derive Eq.(3), non-linearity of the
($\varphi^I_{31}\rightarrow\varphi^V_{31}$) transformation, or
additional dependence in that transformation.

\vspace*{3pt}
A new $(P,\varphi_{31})\rightarrow{\rm [Fe/H]}$ relation was derived
by Nemec \etal (2013) using 26 Kepler-field RRab stars with excellent
light curves. The formula is nonlinear and agrees very well with spectroscopic
iron abundances. Jeon, Ngeow and Nemec (2014) provide the transformation 
from $\varphi^V_{31}$ to $\varphi^{\it Kp}_{31}$, which together with the
relation allows for calculating [Fe/H] from the {\it V}-band light curve:
\begin{subequations}
\begin{equation}
{\setlength\arraycolsep{2pt}
\renewcommand{\arraystretch}{1.3}
\begin{array}{ccr@{.}lcr@{.}lcr@{.}lcr@{.}lcr@{.}l}
{\rm [Fe/H]_N} = &-  & 8&65 &-  & 40&12P&+  & 5&96 \varphi^{\it Kp}_{31} &+  & 6&27\varphi^{\it Kp}_{31} P &-  & 0&72 (\varphi^{\it Kp}_{31})^2 \\
                 &\pm& 4&64 &\pm&  5&18 &\pm& 1&72                       &\pm& 0&96                        &\pm& 0&17 \\
\end{array}}
\end{equation}
\begin{equation}
\varphi^{\it Kp}_{31}=\varphi^V_{31}  + 0.174\qquad(\pm 0.085)
\end{equation}
\end{subequations}

\renewcommand{\arraystretch}{1}
\vspace*{3pt}
The above relation is the most up-to-date and should prove most accurate as
it relies on very high quality and cadence light curves from the Kepler
data for which high-resolution spectroscopic measurements were made with a
derivation of ${\rm [Fe/H]}_{\it spec}$ being a primary goal. It also corrects
the problem of the JK96 formula overestimating [Fe/H] by $\approx0.3~$dex
at low metallicity values. This was partly due to smaller number of stars
at low (${\rm [Fe/H]}<-2.0$~dex) metallicities and inaccurate spectroscopic
metallicity adopted for \eg X~Ari (${\rm [Fe/H]}=-2.10$~dex instead of
${\rm [Fe/H]}\approx-2.65$~dex).  However, the weak point of this method is
Eq.(5b), which has a fairly large error of $\pm0.085$ and was based on only
34 stars, among which there were a few Blazhko RRab.

\vspace*{3pt}
All the above relations are on the High Dispersion Spectroscopy (HDS)
metallicity scale of Jurcsik (1995, hereafter J95).

\vspace*{3pt}
In this paper, we want to calculate an accurate transformation between the
phase parameter $\varphi_{31}$ between the {\it I}- and the {\it V}-band.
We think this is important for at least two reasons:
\begin{itemize}
\item[1.] As one can see, most attempts to calculate the photometric metallicity rely
either on the {\it V}-band light curves, or on a transformation of the phase
parameter from {\it I} (or {\it Kp}) to {\it V}.
\item[2.] The only relation that directly uses the {\it I}-band data is the
one of S05, but as noticed by Deb and Singh (2014), it shows an offset of
0.07~dex as compared to the formula of JK96, after transforming it to the
{\it I}-band. However, the ($\varphi^I_{31}\rightarrow\varphi^V_{31}$)
transformation that was used had been determined with a small sample of only
29 stars, so its reliability may be questioned.
\end{itemize}
 
\Section{OGLE-IV Data}
\subsection{RRab Light Curves}

\vspace*{5pt} For the purpose of this study we used the OGLE-IV RRab sample
that consists of 27\,258 stars in the Galactic bulge (Soszyñski \etal 2014)
and 27\,081 in the LMC (Soszyñski \etal 2016). We do not include the SMC
RRL sample because of smaller number of observations in the {\it V}-band.

To ensure good Fourier fits to the light curves, we selected stars that
have at least 80 observations in the {\it V}-band and at least 300 in the
{\it I}-band. After this selection we were left with 14\,755 objects:
14\,033 in the LMC and 722 in the bulge. An average number of points per
light curve is 640 in the {\it I}-band and 170 in the {\it V}-band.

\vspace*{5pt}
\subsection{Fourier Decomposition of the Light Curves}
\vspace*{3pt}
For each star we perform the Fourier analysis of its light curve, separately
for the {\it V}- and the {\it I}-band in the form:
$$I(t)=A^V_0+\sum\limits_{k=1}^nA^V_k\times\sin(2\pi kx+\varphi^V_k)$$
$$V(t)=A^I_0+\sum\limits_{k=1}^nA^I_k\times\sin(2\pi kx+\varphi^I_k)$$
where $x=({\rm HJD-HJD_0})/P$ and $n=8$ is an order of the fit. We then shift
all phases $\varphi^V_k$ and $\varphi^I_k$ to the $0{-}2\pi$ range.

Next we calculate phase combinations $\varphi^I_{k1}=\varphi^I_k-k\varphi^I_1$
and $\varphi^V_{k1}=\varphi^V_k-k\varphi^V_1$, where $k=2,3,4,5,6$.
According to the prescription of JK96 one should choose $\varphi_{k1}$ such
that it is closest to the mean observed sample average. They provide the
observed averages for phase combinations in the {\it V}-band:\\
$\langle\varphi^V_{21}\rangle=2.4$, $\langle\varphi^V_{31}\rangle=5.1$, $\langle\varphi^V_{41}\rangle=1.6$,
$\langle\varphi^V_{51}\rangle=4.3$, $\langle\varphi^V_{61}\rangle=1.0$\\
and draw attention to the careful choice of $\varphi^V_{31}$ before using
Eq.(1).
Thanks to a large number of stars in our sample, we are
able to determine precise sample averages in the $0-2\pi$ range in
both bands, which mostly agree very well with those of JK96:\\
$\langle\varphi^V_{21}\rangle=2.4$, $\langle\varphi^V_{31}\rangle=5.1$, $\langle\varphi^V_{41}\rangle=1.6$, 
$\langle\varphi^V_{51}\rangle=4.3$, $\langle\varphi^V_{61}\rangle=0.8$,\\
$\langle\varphi^I_{21}\rangle=2.8$, $\langle\varphi^I_{31}\rangle=5.8$, $\langle\varphi^I_{41}\rangle=2.6$, 
$\langle\varphi^I_{51}\rangle=5.6$, $\langle\varphi^I_{61}\rangle=2.2$.\\
We then shift all $\varphi^V_{k1}$ and $\varphi^I_{k1}$ values within
$\pm\pi$ of the sample averages.

\vspace*{5pt}
\subsection{The Sample Cleaning}
\vspace*{3pt}
The median scatter of observations around a Fourier model is 0.05~mag with
a standard deviation of 0.02~mag in the {\it I}-band, and 0.05~mag with a
standard deviation of 0.04~mag in the {\it V}-band. This scatter is mostly
caused by Blazhko RRab stars, which constitute about 20\% of all RRab stars
in the LMC (Soszyñski \etal 2009) and about 30\% of all RRab in the
Galactic bulge (Soszyñski \etal 2011). According to recent Kepler and
highest-quality ground based data, as many as 50\% of RRab are expected to
show the Blazhko effect (\eg Benk\H{o} \etal 2010). However, by applying
the 50\% cut to our data, we would risk removing non-modulated stars with
higher light curve scatter, \eg faint stars that might be indistinguishable
from Blazhko variables, and bias the sample. In order to clean our sample
from Blazhko stars we reject objects based primarily on their {\it I}-band
scatter, because there are 3--4 times as many observations in the {\it
I}-band, making the detection of the true Blazhko effect much more
secure. Thus we exclude RRab that have an {\it I}-band scatter larger than
0.07~mag, which removes $\approx20\%$ of stars from our sample. We also
exclude RRab variables that have a {\it V}-band scatter larger than
0.09~mag which leaves us with 11476 RRab. In order to further increase
sample purity we also reject the faintest RRab stars with $I>19$~mag and
$V>19.6$~mag which leaves 9527 RRab in the sample.

\vspace*{3pt}
JK96 introduce a deviation parameter $D_F$ for measuring the relative accuracy
of the [Fe/H] prediction:
$$D_F=|F_{\rm obs}-F_{\rm calc}|/\sigma_F,$$
where $F_{\rm obs}$ is the observed value of a given Fourier parameter and
$F_{\rm calc}$ is its predicted value based on other observed parameters and 
$\sigma_F$ is the respective standard deviation. The relations used to
calculate $F_{\rm calc}$ are provided in Table~6 of JK96. If the maximum
deviation parameter for a given star $D_{\rm m}=\max\lbrace D_F\rbrace$ is
small ($D_{\rm m}<3$) then the light curve satisfies the compatibility
condition and can be used for calculating the star's metallicity.
\begin{figure}[htb]
\includegraphics[height=4.1cm]{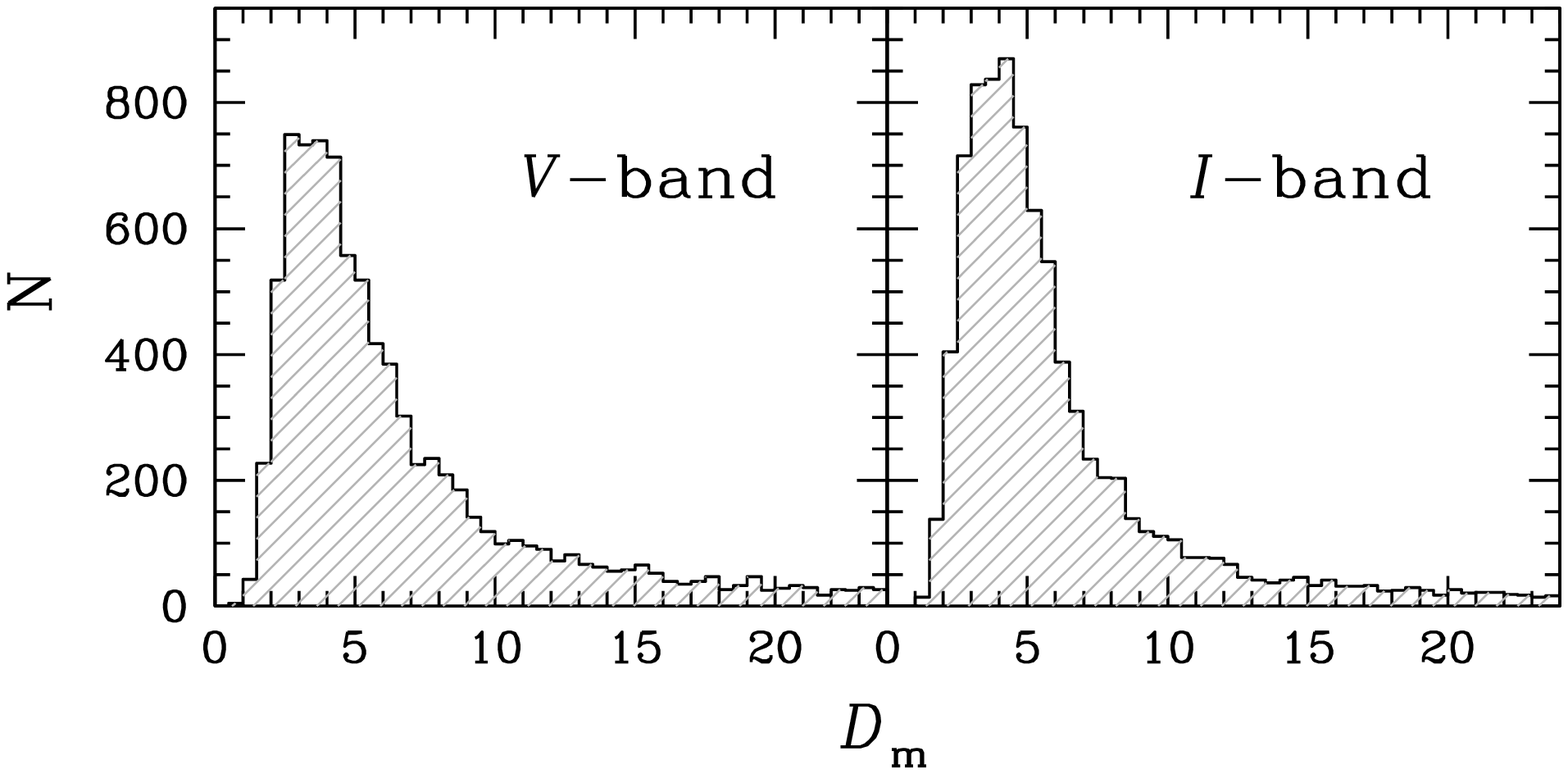}\hfil\includegraphics[height=4.1cm]{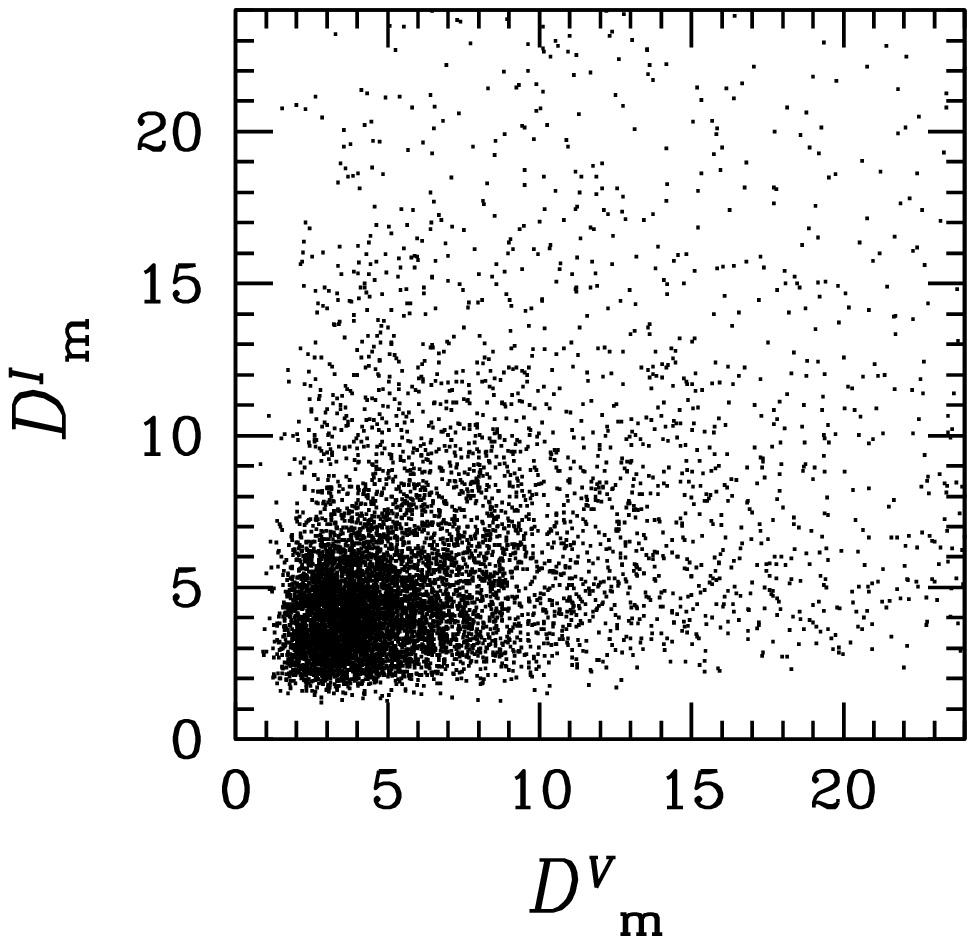}
\vskip5pt
\FigCap{Maximum deviation parameter distribution for a cleaned sample
of 9527 RRab, calculated from the {\it V}-band data ({\it left panel}) and the
{\it I}-band data ({\it middle panel}). {\it Right panel} shows the relation
between the two deviation parameters.}
\end{figure}

Fig.~1 shows the distribution of $D_{\rm m}$ calculated from the {\it V}-band
data (left panel) and the {\it I}-band data (middle panel). Only 16\% of stars
satisfy the condition $D_{\rm m}<3$ in the {\it V}-band, only 13\% in the
{\it I}-band, and barely 4\% in both bands. It has been previously noticed
(Szczygie³ \etal 2009, Deb and Singh 2010) that the compatibility condition
is indeed very strict and rejects many normal-looking light curves.
Interestingly, there is also no correlation between $D^V_{\rm m}$ and
$D^I_{\rm m}$, as shown in the right panel of Fig.~1, suggesting that
$D_{\rm m}<3$ may not be the best criterion for ''compatible light curve''
selection. This seems probable, as the calibrating sample of JK96 used to
derive parameter interrelations and the compatibility condition, had only 10
objects. Nevertheless, in order to avoid abnormal light curves, rather than
select the ideal ones, we reject stars that have $D_{\rm m}>10$ in either band,
that is 3089 (32\%) stars in our sample, leaving us with 6438 RRab variables.

\Section{Photometric Metallicity Calculation}
\subsection{Relations between Phase Parameters $\varphi^V_{k1}$ and $\varphi^I_{k1}$}
In order to find the relations between phase parameters $\varphi^V_{k1}$
and $\varphi^I_{k1}$, we first bin the data in steps of 0.05, 0.08, 0.1,
0.15, and 0.2 for $k=2,3,4,5$, and 6 respectively, both in the {\it I}- and
{\it V}-band, requiring at least 40 points in the bin. Fig.~2 shows the
data, where small gray dots mark all 6438 RRab and large gray dots
represent the binned data. The dotted lines show the $1\sigma$ and
$3\sigma$ range within the median. Next we fit a second order polynomial to
the binned data in the form:
$$y=ax^2+bx+c$$
where $x=\varphi^I_{k1}$ and $y=\varphi^V_{k1}$. 
The errors of the fit parameters ($\sigma_a$, $\sigma_b$, $\sigma_c$)
are calculated using the bootstrap method for $N=10\,000$ bootstrap samples.
The final parameters of the fit are listed in Table~1 and in Fig.~2, where
the red solid line represents the final fit.

The choice of the second order polynomial was preceded by a series of tests
where other functions were fitted to the data. By comparing the $\chi^2$ of
the fits we were able to select the best-fitting function. Thanks to the
large number of points we see that the relation is quadratic, rather than
linear as found by Deb and Singh (2010), who had a sample of only 29 RRab
stars.  For comparison, their linear relation is plotted with a dashed line
in the top right panel of Fig.~2.

\renewcommand\arraystretch{1.3}
\MakeTable{|c|r|r|r|r|}{12.5cm}{Fit parameters for $y=ax^2+bx+c$}
{\hline
 & \multicolumn{1}{|c|}{a} & \multicolumn{1}{|c|}{b} & \multicolumn{1}{|c|}{c} & \multicolumn{1}{|c|}{$\sigma_{\rm fit}$} \\
\hline
$\varphi^I_{21} \rightarrow \varphi^V_{21}$  & $0.405 \pm 0.033$ & $-1.646 \pm 0.179$ & $ 3.815 \pm 0.245$ & 0.001 \\
$\varphi^I_{31} \rightarrow \varphi^V_{31}$  & $0.122 \pm 0.017$ & $-0.750 \pm 0.187$ & $ 5.331 \pm 0.523$ & 0.004 \\
$\varphi^I_{41} \rightarrow \varphi^V_{41}$  & $0.082 \pm 0.013$ & $0.247  \pm 0.061$ & $ 0.401 \pm 0.070$  & 0.009 \\
$\varphi^I_{51} \rightarrow \varphi^V_{51}$  & $0.047 \pm 0.010$ & $0.141  \pm 0.112$ & $ 2.075 \pm 0.307$  & 0.019 \\
$\varphi^I_{61} \rightarrow \varphi^V_{61}$  & $0.057 \pm 0.011$ & $0.364  \pm 0.052$ & $-0.357 \pm 0.051$ & 0.100 \\
\hline
}
\renewcommand\arraystretch{1.0}

\begin{figure}[p]
\includegraphics[height=5.9cm]{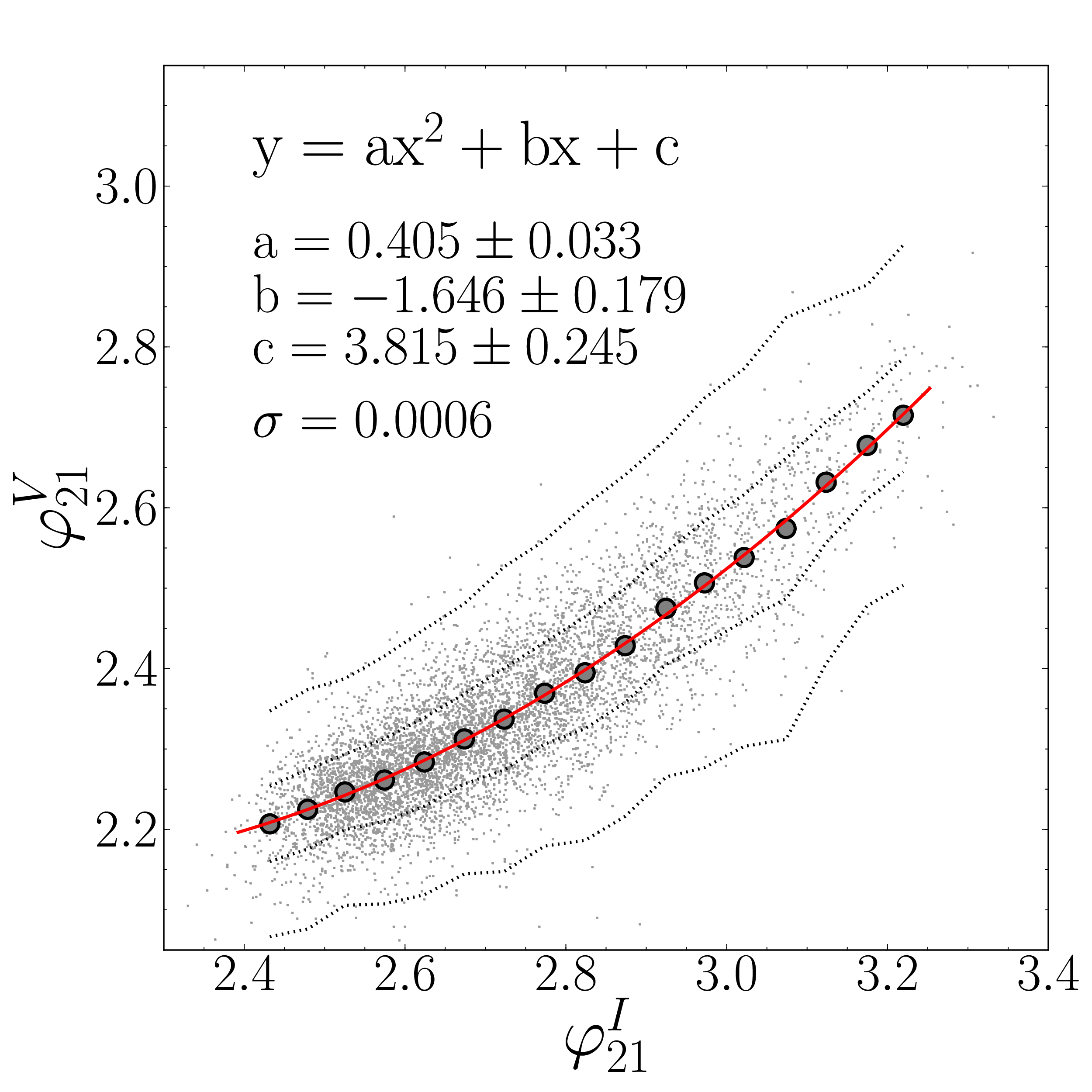} \hfil \includegraphics[height=5.9cm]{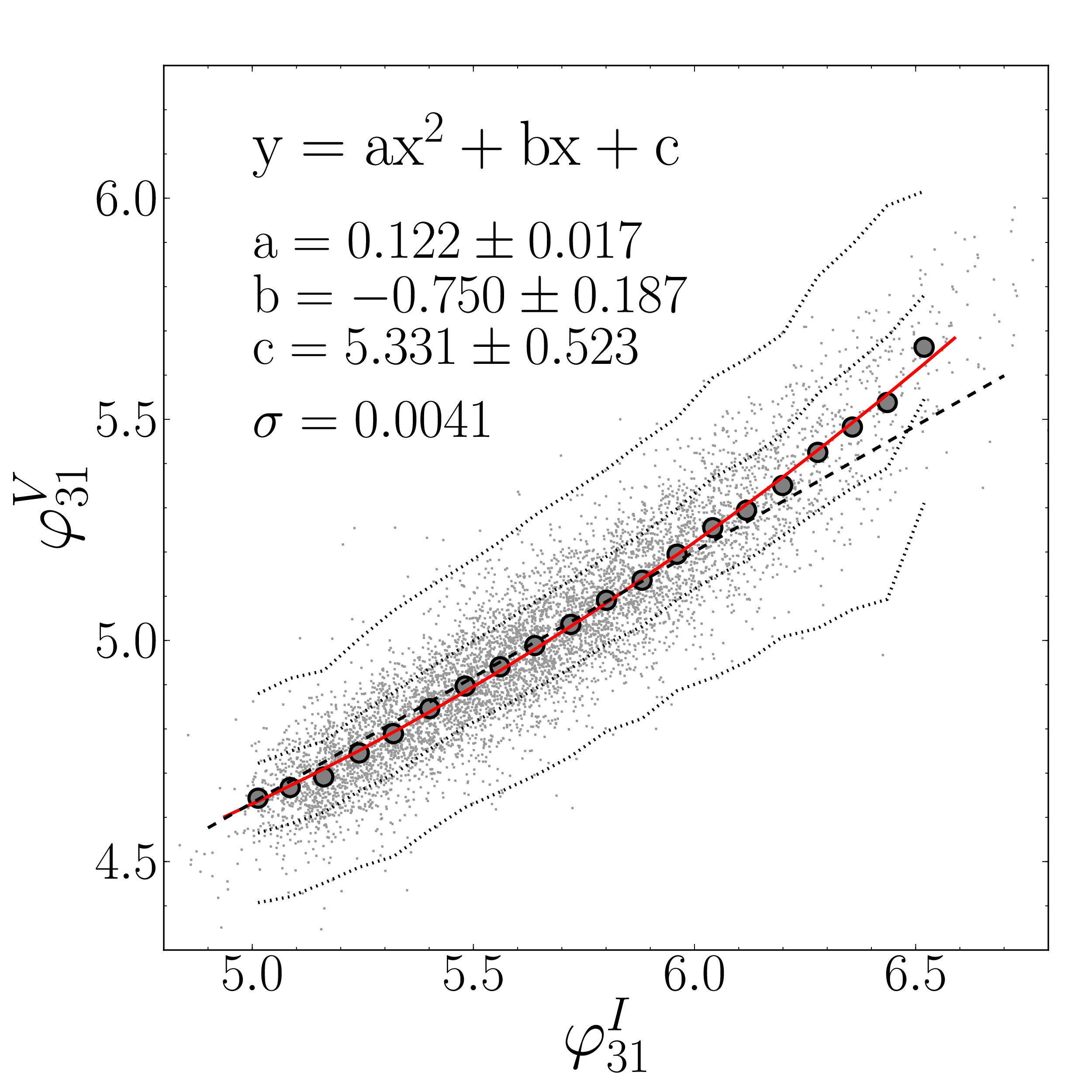} \\
\includegraphics[height=5.9cm]{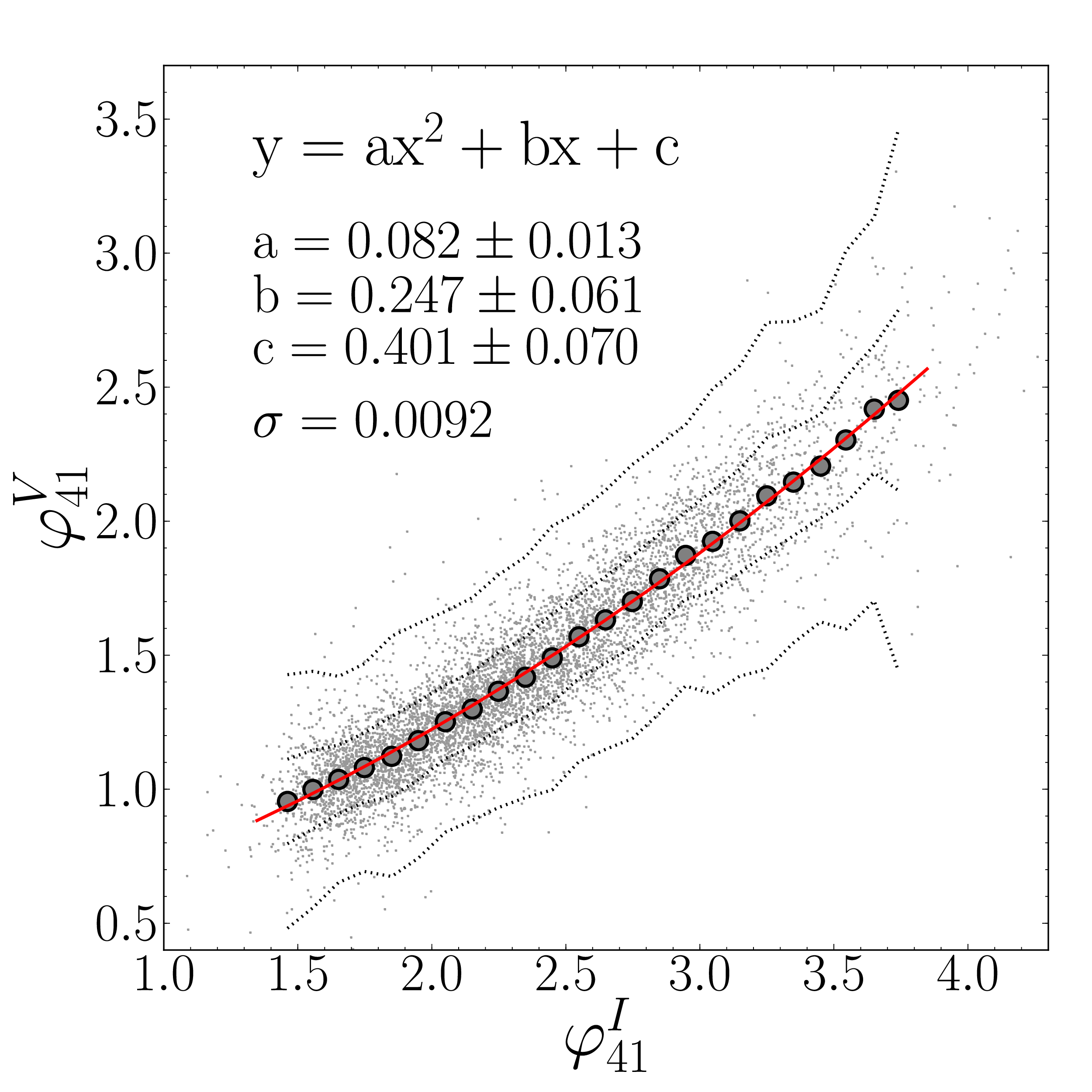} \hfil \includegraphics[height=5.9cm]{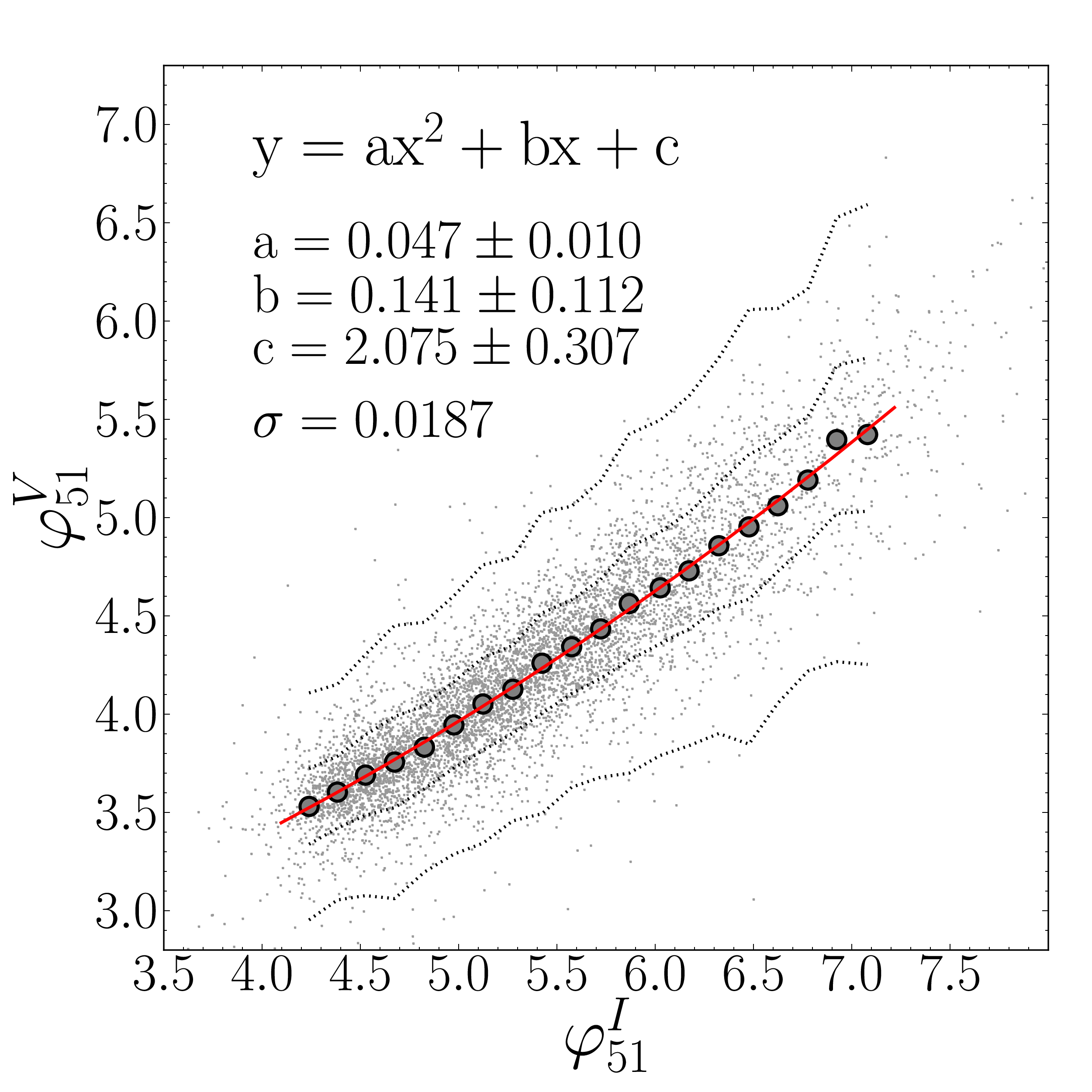} \\
\includegraphics[height=5.9cm]{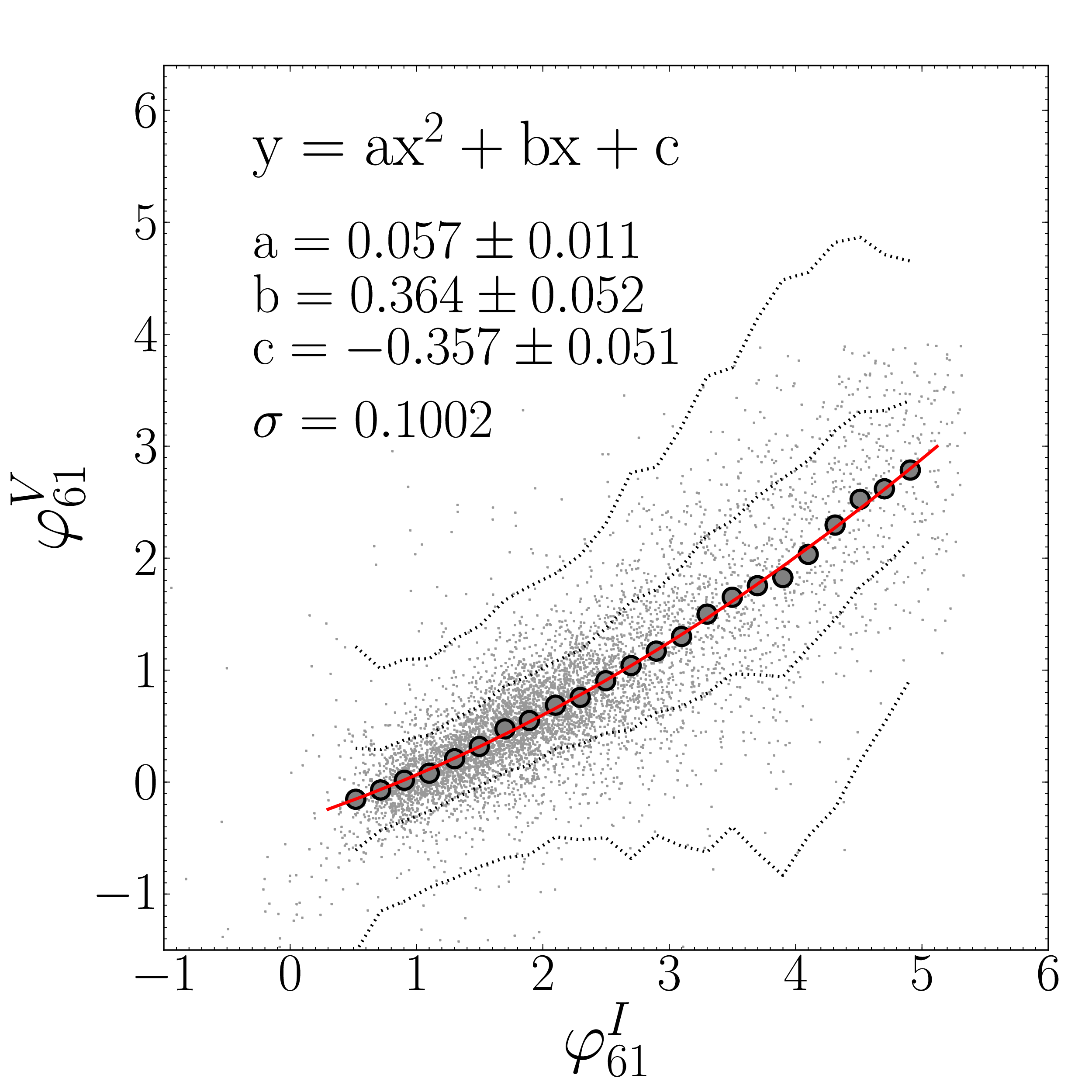}
\vskip5pt
\FigCap{Interrelations between phase parameters $\varphi^I_{k1}$ and
  $\varphi^V_{k1}$. The red solid line is the second order polynomial fit
  to the large dots that represent the median value within a given bin.
  Fit parameters are given in Table~1 and in {\it all panels}. The
  dotted black lines mark $1\sigma$ and $3\sigma$ deviations from the
  median value. The dashed line in the {\it top right panel} shows the linear
  relation from Deb and Singh (2010). Note that the phase parameters come
  from a sine series fit to the light curves.}
\end{figure}

\subsection{Photometric Metallicity Formula in the I-band}
We use the interrelation ($\varphi^I_{31} \rightarrow \varphi^V_{31}$)
estimated in Section~4.3:
$$\varphi^V_{31}=0.122(\varphi^I_{31})^2-0.750\varphi^I_{31}+5.331\eqno(6)$$
and paste it into the relation of JK96 (Eq.~1),
thus obtaining the formula to calculate ${\rm [Fe/H]}_{\it JK}$
based on the {\it I}-band light curve:
$${\rm [Fe/H]^I}_{\it JK}=2.132-5.394P-1.009\varphi^I_{31}+0.164(\varphi^I_{31})^2\eqno(7)$$
In order to see how accurate the calculation of the iron abundance is, in
Fig.~3 we compare ${\rm [Fe/H]}_{\it JK}$ (based on the {\it V}-band light
curve, Eq.~1) with ${\rm [Fe/H]}^I_{\it JK}$ (based on the {\it I}-band light
curve, Eq.~7). We see that the differences can be very large, up to
0.4~dex, with a $\sigma$ of 0.13~dex, which was expected when looking at
the scatter in Fig.~2. This effect does not disappear if a more strict
sample of the highest quality light curves is chosen (see the gray
histogram in the left panel of Fig.~3. and black points in the right
panel). Moreover, there is a clear correlation in the right panel of
Fig.~3, which means that the ($\varphi^I_{31}\rightarrow \varphi^V_{31}$)
interrelation may depend on metallicity.
\begin{figure}[htb]
\includegraphics[width=5.7cm]{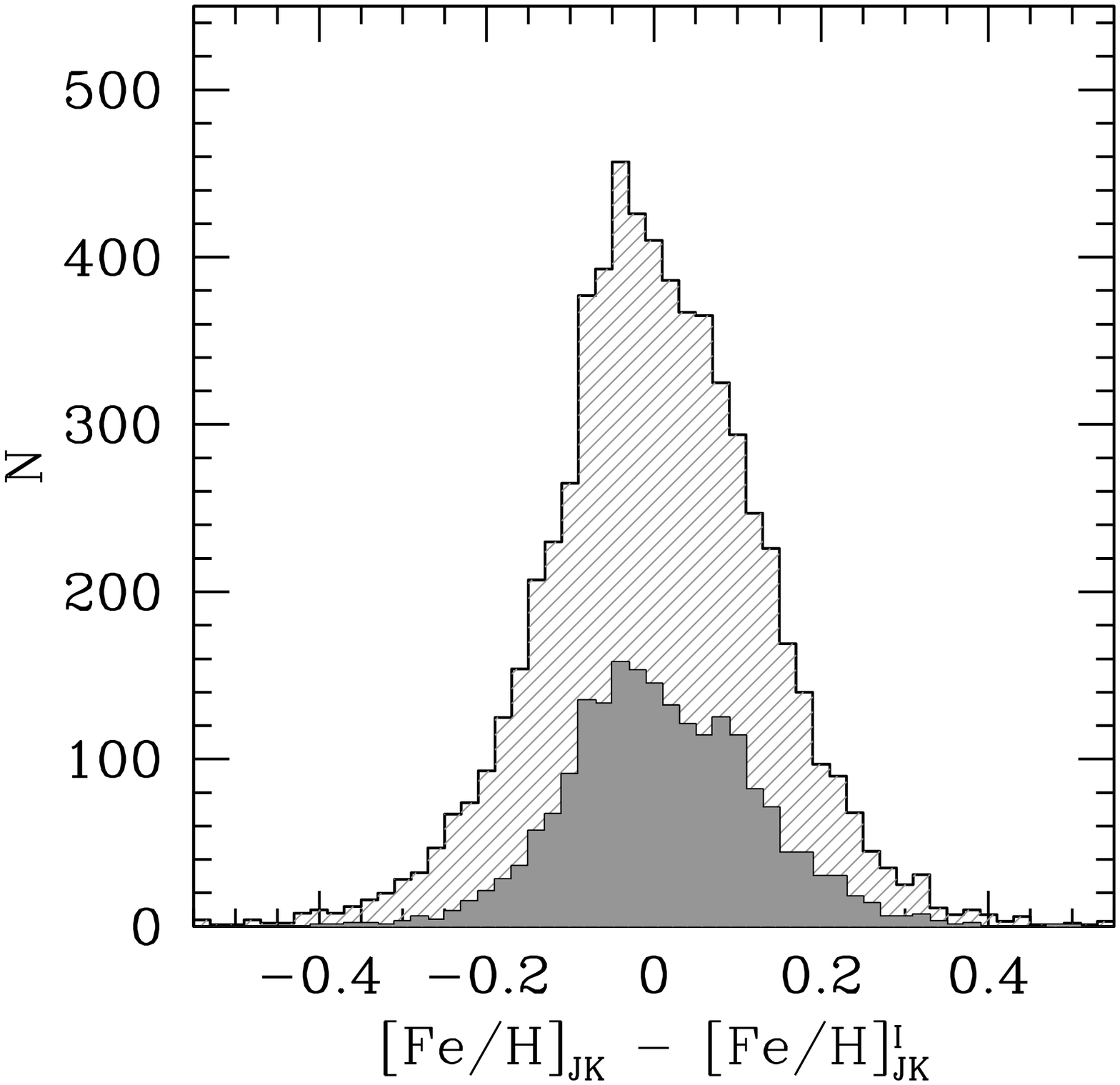} \hfil \includegraphics[width=5.7cm]{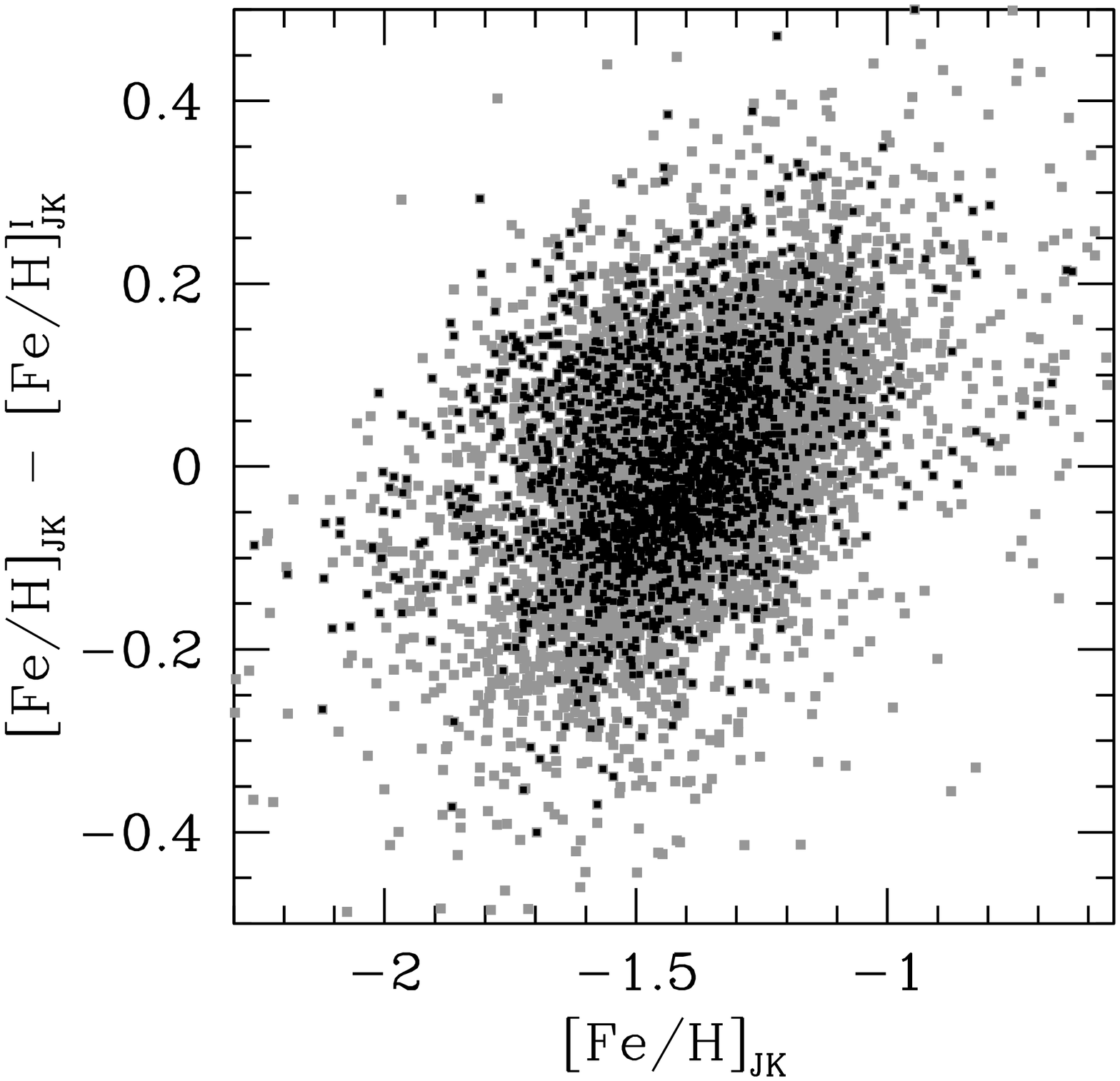}
\vskip5pt
\FigCap{{\it Left panel:} histogram of metallicity differences
${\rm [Fe/H]}_{\it JK}-{\rm [Fe/H]}^I_{\it JK}$. Dashed area represents all
6438 RRab, while the gray filled area a subset -- 2034 RRab with highest
quality light curves. {\it Right panel:} metallicity difference
${\rm [Fe/H]}_{\it JK}-{\rm [Fe/H]}^I_{\it JK}$ against ${\rm [Fe/H]}_{\it JK}$
for the full sample (gray points) and for the high-quality sample (black points).}
\end{figure}

In order to verify if there is a metallicity dependence of the
($\varphi^I_{31}\rightarrow\varphi^V_{31}$) interrelation we repeat the
fitting procedure described in Section~4.1, but this time for several
metallicity bins. We first calculate the ${\rm [Fe/H]}_{\it JK}$ photometric
metallicity using the {\it V}-band light curve phase parameter. We then
separate stars into eight metallicity bins chosen to ensure fair resolution
and a sufficient number of stars in the bin, as presented in Table~2. The
fitting results are shown in Fig.~4, where we plot $\varphi^V_{31}$ \vs
$\varphi^I_{31}$ similarly as in Fig.~2, and each of the eight fits,
represented with a different color line. We see a clear trend in
($\varphi^I_{31}\rightarrow\varphi^V_{31}$) dependence from the most to the
least metal poor RRab, especially for $\varphi^I_{31}<6$. Individual fit
parameters are presented in Table~2.

\def\arraystretch{1.3}
\MakeTableee{|c|c|r|r|r|r|r|}{12.5cm}{Metallicity bin characteristics and fit parameters}
{\hline
\multicolumn{2}{|c|}{${\rm [Fe/H]}_{\it JK}$ [dex]} & \multirow{2}{*}{${N_{\rm stars}}$} & \multicolumn{4}{|c|}{Fit parameters for $\varphi^V_{31}={\rm a}(\varphi^I_{31})^2+{\rm b}\varphi^I_{31}+{\rm c}$}\\
\cline{1-2}
\cline{4-7}
range & median & & \multicolumn{1}{|c|}{a} & \multicolumn{1}{|c|}{b} & \multicolumn{1}{|c|}{c} & \multicolumn{1}{|c|}{$\sigma_{\rm fit}$} \\
\hline
$\leq -1.7$    & $-1.80$ &  508 & $ 0.098 \pm 0.065 $ & $-0.397 \pm 0.733 $ & $ 4.043 \pm 2.053 $ & 0.003 \\
$(-1.7; -1.6]$ & $-1.64$ &  526 & $ 0.049 \pm 0.060 $ & $ 0.104 \pm 0.682 $ & $ 2.792 \pm 1.919 $ & 0.004 \\
$(-1.6; -1.5]$ & $-1.55$ &  979 & $ 0.161 \pm 0.049 $ & $-1.164 \pm 0.549 $ & $ 6.383 \pm 1.530 $ & 0.003 \\
$(-1.5; -1.4]$ & $-1.45$ & 1251 & $ 0.023 \pm 0.067 $ & $ 0.330 \pm 0.754 $ & $ 2.356 \pm 2.097 $ & 0.003 \\
$(-1.4; -1.3]$ & $-1.35$ & 1238 & $ 0.141 \pm 0.041 $ & $-0.988 \pm 0.459 $ & $ 6.072 \pm 1.275 $ & 0.008 \\
$(-1.3; -1.2]$ & $-1.26$ &  962 & $ 0.091 \pm 0.060 $ & $-0.440 \pm 0.667 $ & $ 4.600 \pm 1.853 $ & 0.006 \\
$(-1.2; -1.1]$ & $-1.16$ &  506 & $-0.079 \pm 0.045 $ & $ 1.464 \pm 0.504 $ & $-0.660 \pm 1.416 $ & 0.001 \\
$> -1.1$       & $-1.00$ &  468 & $ 0.032 \pm 0.040 $ & $ 0.218 \pm 0.461 $ & $ 2.858 \pm 1.337 $ & 0.002 \\
\hline
}
\renewcommand\arraystretch{1.0}

\vspace*{-11mm}
\begin{figure}[htb]
\begin{center}
\includegraphics[height=7.5cm]{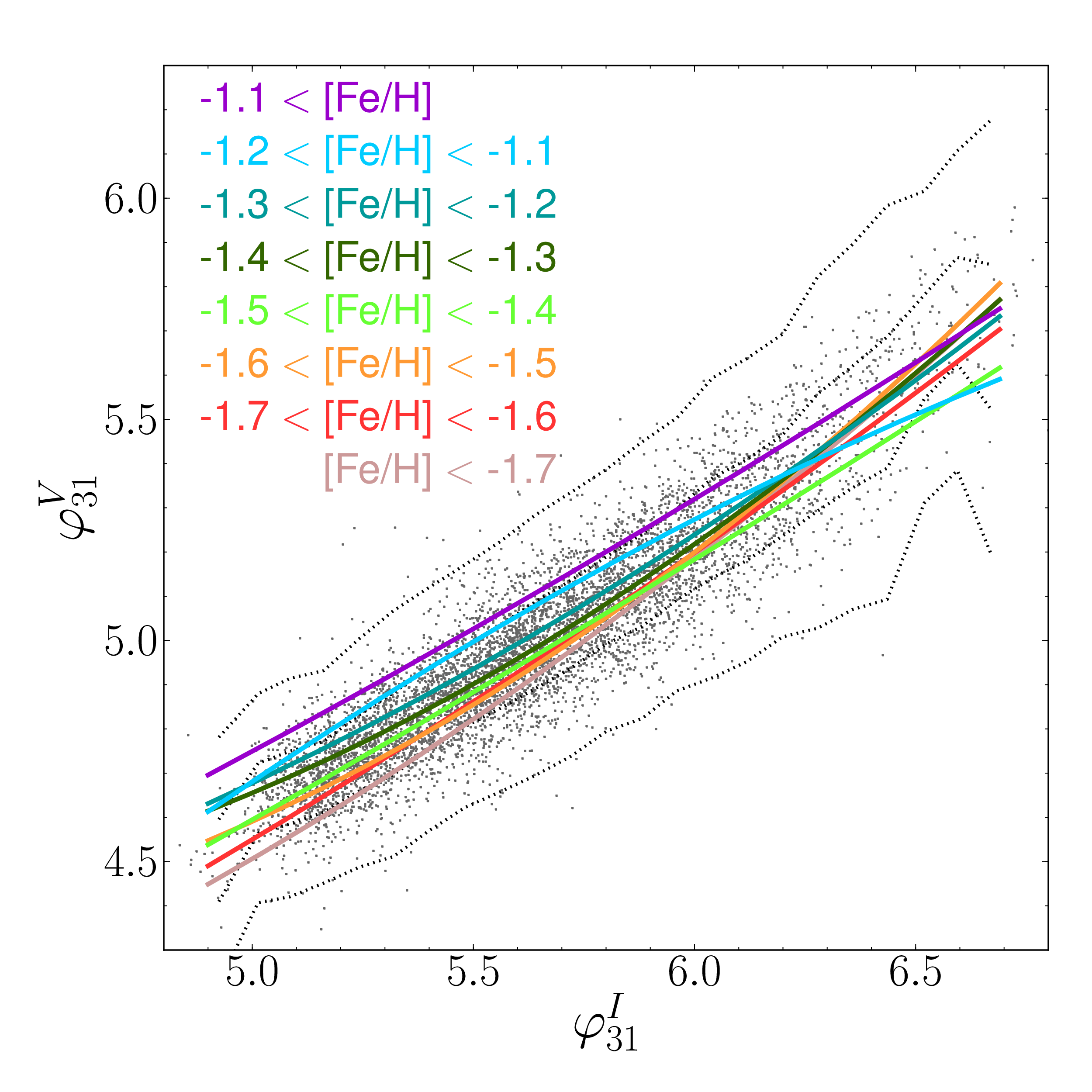}
\end{center}
\vskip-17pt\FigCap{Interrelation between phase parameters
  $\varphi^I_{31}$ and $\varphi^V_{31}$. Dots represent the same sample of
  RRab as in Fig.~2 and the dotted black lines mark $1\sigma$ and $3\sigma$
  deviations from the median value. Different color lines mark second order
  polynomial fits within different JK96 metallicity bins, as described in
  the legend. Fit parameters are given in Table~2.}
\end{figure}

We then calculate ${\rm [Fe/H]}^I_{N1}$ for each star using the relation of
N13 (Eq.~5) and the ($\varphi^I_{31} \rightarrow \varphi^V_{31}$)
transformation from Eq.(6). The resulting formula becomes quite complicated
with eight elements and $\varphi^I_{31}$ to the fourth power, so we will
refrain from providing it in that form, also because repeated rounding of
the coefficients lowers the final accuracy.  We also calculate ${\rm
[Fe/H]}^I_{N2}$ using the relation of N13 (Eq.~5) and the
($\varphi^I_{31} \rightarrow \varphi^V_{31}$) transformation, but this time
the transformation equation coefficients are taken from Table~2, \ie they
depend on the stars metallicity. The results are presented in Fig.~5. The
left panel shows how metallicities calculated from metallicity dependent
and independent formulae change with metallicity, which was expected from
Figs.~3 and~4. The histogram of these differences (${\rm [Fe/H]}^I_{N1}-
{\rm [Fe/H]}^I_{N2}$) is presented in the middle panel of Fig.~5. The ${\rm
[Fe/H]}^I_{N1}-{\rm [Fe/H]}^I_{N2}$ differences for individual stars
reach 0.3~dex, but are typically lower than 0.1~dex. Finally, the right
panel shows a histogram of photometric metallicities ${\rm [Fe/H]}^I_{N1}$
and ${\rm [Fe/H]}^I_{N2}$. Both distributions have a similar shape,
although the metallicity dependent formula produces a slightly wider
profile. Nevertheless, the overall similarity of the two distributions is
reassuring -- even though we are not able to obtain a single relation
between $\varphi^I_{31}$ and $\varphi^V_{31}$ because it depends on
metallicity, the difference between results, \ie whether accounting for
that dependence or not, is not very large. As stressed before, this is true
only when statistically analyzing large samples of RRab, not in the case of
star-to-star comparisons.

\vspace*{-7pt}
\begin{figure}[b]
\includegraphics[width=3.9cm]{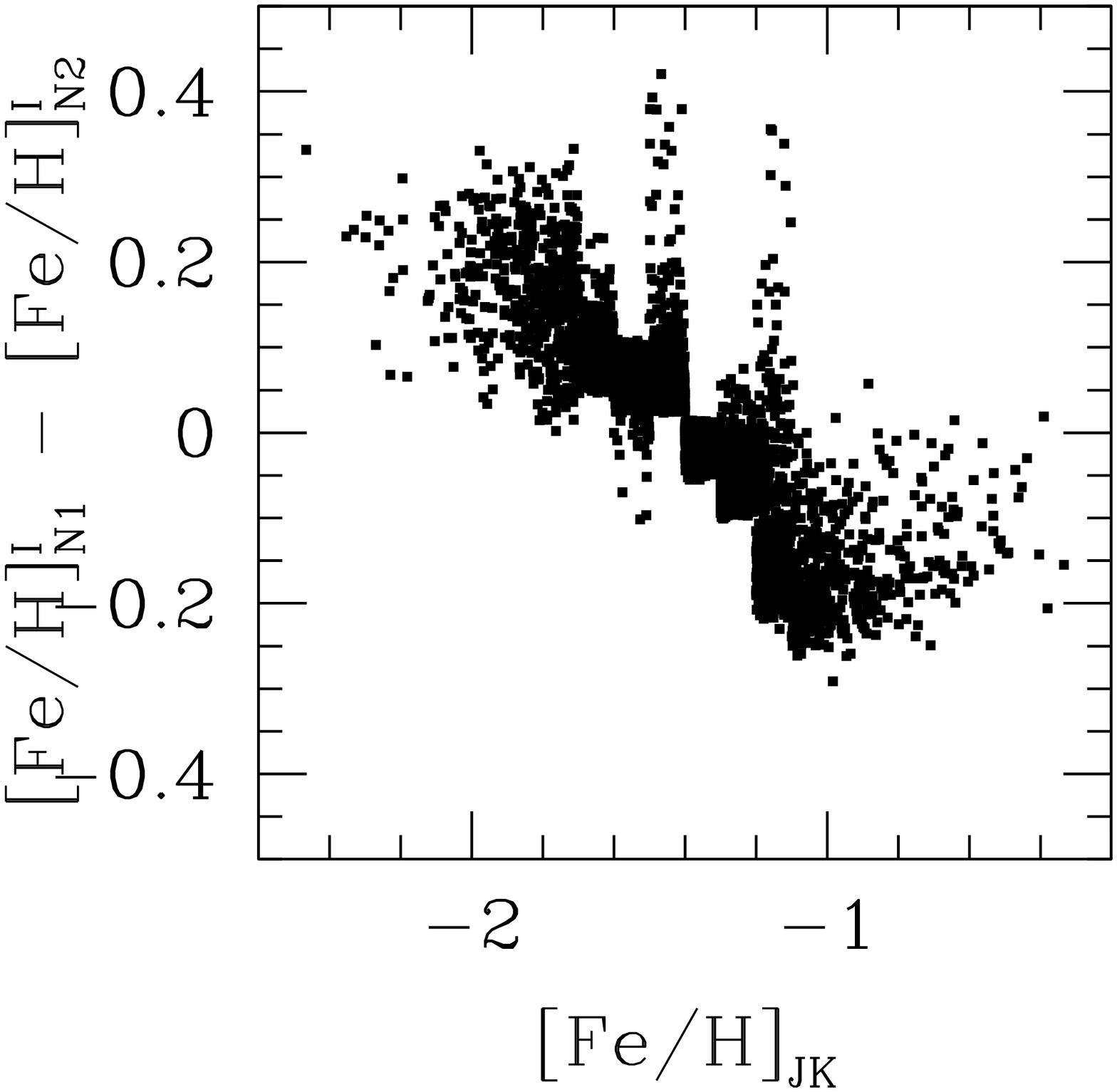}\hskip3mm \includegraphics[width=3.9cm]{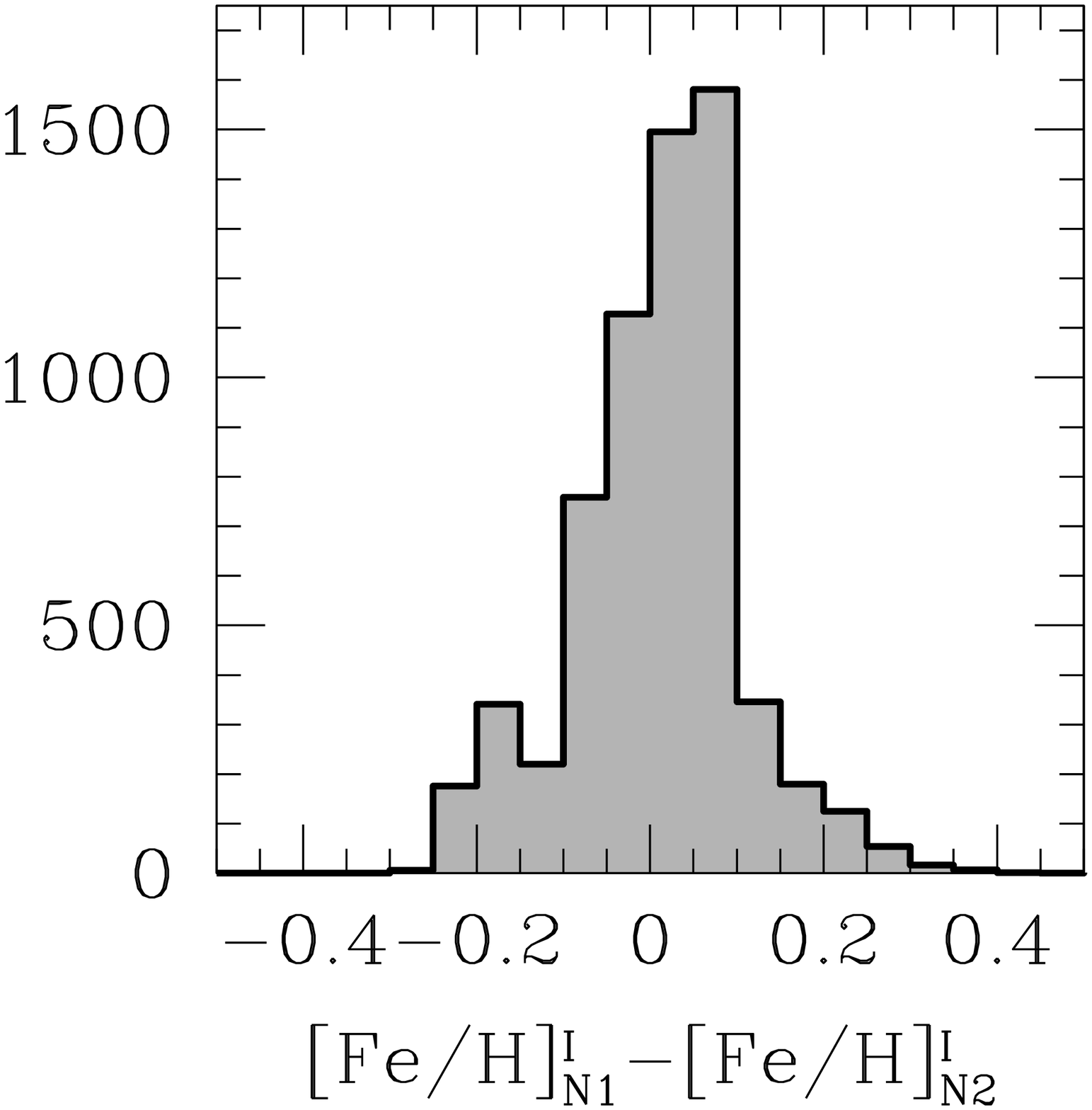}\hskip3mm \includegraphics[width=3.9cm]{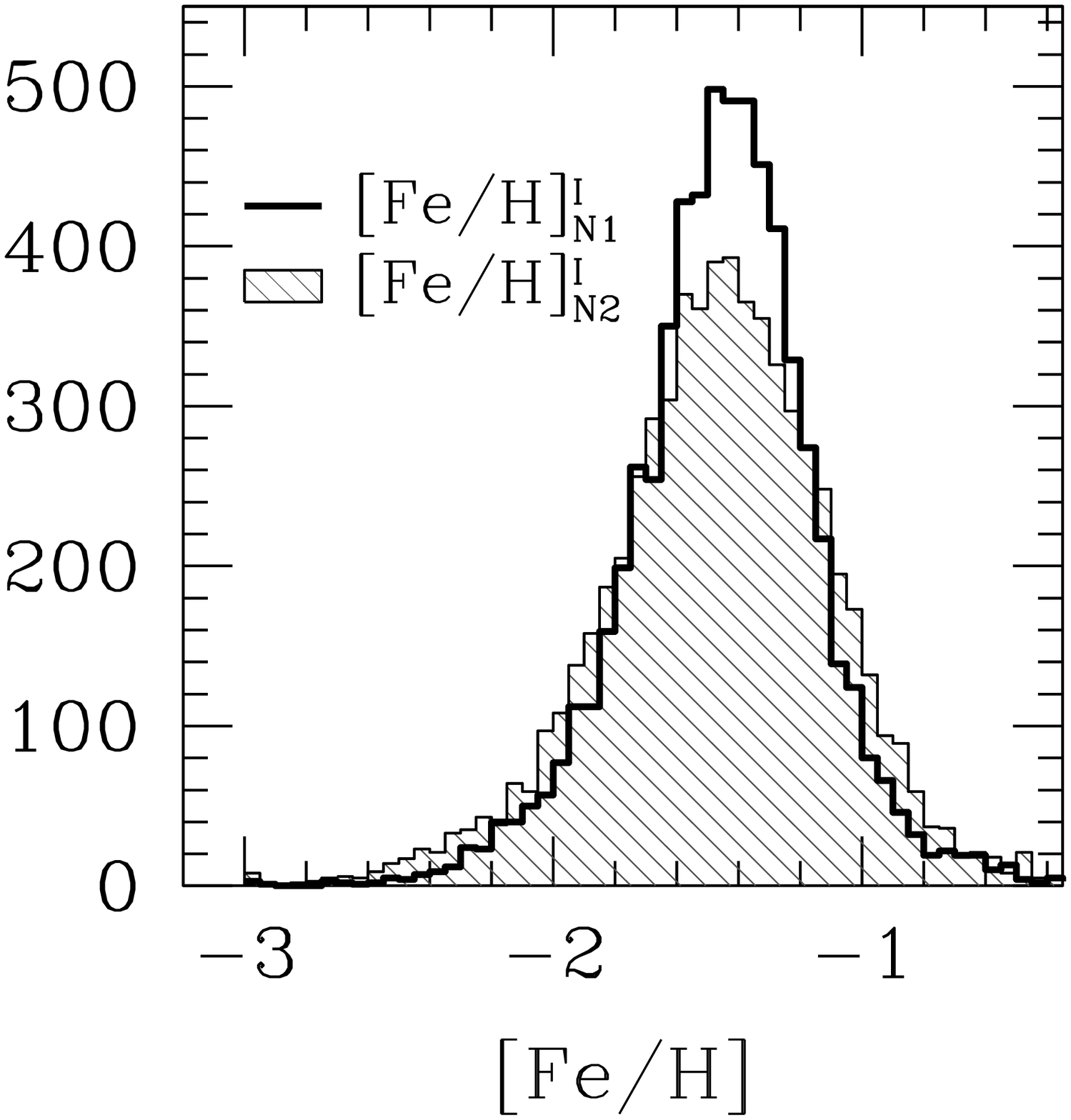} \\
\vskip-5pt
\FigCap{Comparison of photometric metallicities calculated using
metallicity independent (${\rm [Fe/H]}_{N1}$) and metallicity dependent
(${\rm [Fe/H]}_{N2}$) formulae. {\it Left panel:} metallicity difference
${\rm [Fe/H]}_{N1}-{\rm [Fe/H]}_{N2}$ \vs ${\rm [Fe/H]}_{\it JK}$. {\it Middle
panel:} a histogram of metallicity differences ${\rm
[Fe/H]}_{N1}-{\rm [Fe/H]}_{N2}$. {\it Right panel:} a histogram of ${\rm
[Fe/H]}_{N1}$ and ${\rm [Fe/H]}_{N2}$.}
\vskip3mm
\includegraphics[width=5.5cm]{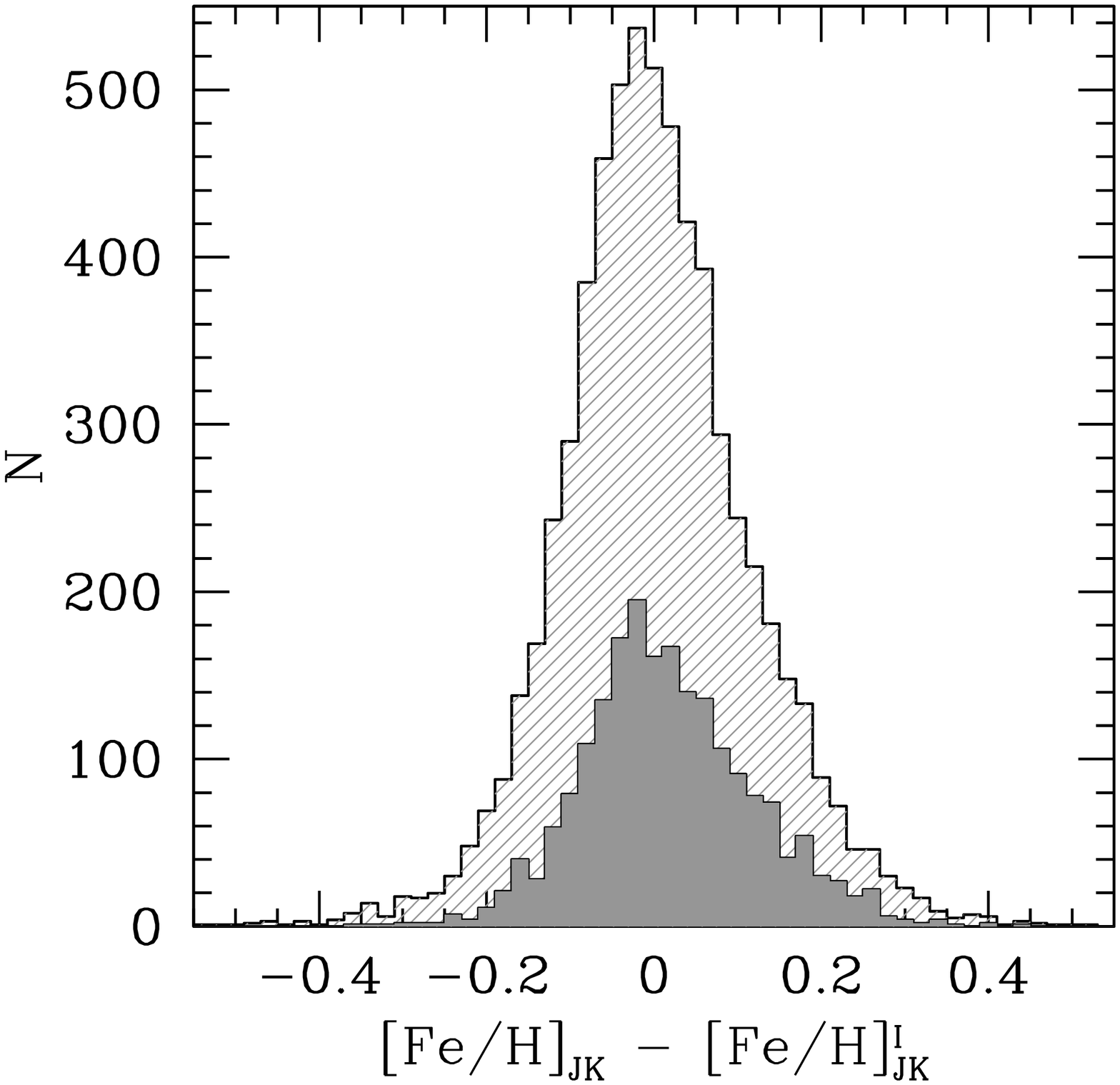} \hfil \includegraphics[width=5.5cm]{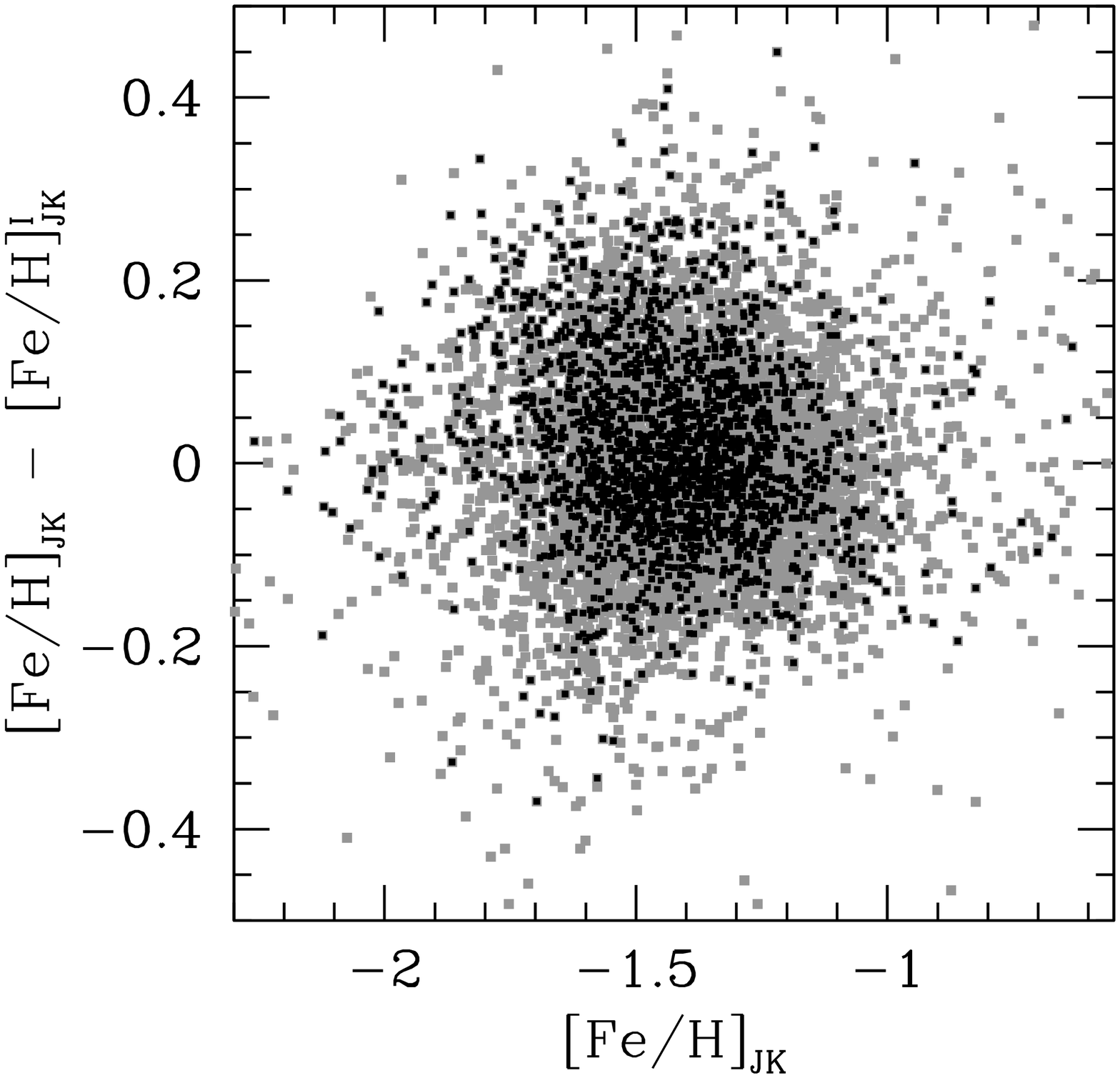}
\vskip3pt
\FigCap{Analog of Fig.~3, only this time the metallicity dependence of the
($\varphi^I_{31} \rightarrow \varphi^V_{31}$) interrelation is taken into
account. {\it Left panel:} histogram of metallicity differences
${\rm [Fe/H]}_{\it JK}-{\rm [Fe/H]}^I_{\it JK}$. Dashed area represents all
6438 RRab, while the gray filled area a subset -- 2034 RRab with highest
quality light curves. {\it Right panel:} metallicity difference
${\rm [Fe/H]}_{\it JK}-{\rm [Fe/H]}^I_{\it JK}$ against ${\rm [Fe/H]}_{\it JK}$
for the full sample (gray points) and for the high-quality sample (black points).}
\end{figure}

Fig.~6 is an analog of Fig.~3, \ie it compares ${\rm [Fe/H]}_{\it JK}$ (based
on the {\it V}-band light curve, Eq.~1) with ${\rm [Fe/H]}^I_{\it JK}$ (based
on the {\it I}-band light curve). Only this time we are not using a single
Eq.(6) to transform $\varphi^I_{31}$ to $\varphi^V_{31}$, but a set of
equations from Table~2, which are then pasted into Eq.(1). We see, that
after accounting for the metallicity dependence of the
($\varphi^I_{31}\rightarrow \varphi^V_{31}$) interrelation, the correlation
visible in Fig.~3 vanishes in Fig.~6, although the scatter remains high,
suggesting that this is an intrinsic property of this interrelation.

\subsection{Applicability of the ($\varphi^I_{31} \rightarrow \varphi^V_{31}$) interrelation}
As shown in this section, the applicability of the ($\varphi^I_{31}
\rightarrow\varphi^V_{31}$) transformation is limited. Since it depends
on metallicity, there is a set of equations instead of one, and deciding
which of those equations should be used, requires knowing the ${\rm
[Fe/H]}_{\it JK}$ metallicity a priori. However, the fact that one needs to
use the ($\varphi^I_{31}\rightarrow\varphi^V_{31}$) transformation means
that the ${\rm [Fe/H]}_{\it JK}$ calculation is not possible, as there is no
{\it V}-band data available. In such case, one idea is to iterate the whole
process, until the most accurate photometric metallicity is calculated.
Another, simpler, idea is to use the mean relation between $\varphi^I_{31}$
and $\varphi^V_{31}$, \ie Eq.(6). As shown in Fig.~5, such distribution will
be still overestimated at lowest [Fe/H] and underestimated at highest
[Fe/H] (see left panel), but for the majority of stars the differences will
be $<0.1$~dex (see middle panel). In the case of a large sample this effect
does not change the overall distribution significantly -- median and
standard deviations of the two histograms in the right panel of Fig.~5 are:
$\langle {\rm [Fe/H]}^I_{N1}\rangle=-1.44\pm0.31$~dex, and $\langle
{\rm [Fe/H]}^I_{N2}\rangle=-1.45\pm0.38$~dex.

\subsection{Comparison of [Fe/H] from Three Methods}
We calculate photometric metallicities for 6438 RRab using three methods:
${\rm [Fe/H]}_S$ (Eq.~3), ${\rm [Fe/H]}^I_{\it JK}$ (Eq.~1 after transforming
$\varphi^I_{31}$ to $\varphi^V_{31}$), and ${\rm [Fe/H]}^I_{N2}$ (Eq.~5
after transforming $\varphi^I_{31}$ to $\varphi^V_{31}$). The
transformations in the latter two were calculated based on parameters from
Table~2. The left panel of Fig.~7 compares the results and shows
metallicity distributions for the three formulae.  Median and standard
deviation values are as follows:
$\langle{\rm [Fe/H]}_S\rangle=-1.29\pm0.19$~dex,
$\langle{\rm [Fe/H]}^I_{\it JK}\rangle=-1.40\pm0.25$~dex, and
$\langle{\rm [Fe/H]}^I_{N2}\rangle=-1.45\pm0.38$~dex.
All presented distributions are different -- the S05 method produces
highest metallicity values with the lowest dispersion, the transformed
JK96 method gives lower metallicity values and the higher dispersion.
${\rm [Fe/H]}^I_{N2}$ is not only shifted toward lower metallicities, but also
has a much flatter and wider profile. We checked that this is not an effect of error
propagation during ($\varphi^I_{31}\rightarrow\varphi^V_{31}$) and
($\varphi^V_{31}\rightarrow\varphi^{\it Kp}_{31}$) transformations and it is
actually expected after compensating for ${\rm [Fe/H]}^I_{\it JK}$ overestimation
at low metallicity values.
The difference between ${\rm [Fe/H]}^I_{\it JK}$ and ${\rm [Fe/H]}^I_{N2}$ is
further pictured in the right panel of Fig.~7. The overall agreement between
the two methods is rather poor, especially at the low metallicity end.
The same is true, when instead of ${\rm [Fe/H]}^I_{N2}$ we use
${\rm [Fe/H]}^I_{N1}$, \ie the metallicity independent formula -- the
distribution of ${\rm [Fe/H]}^I_{N1}$ is only slightly different from 
that of ${\rm [Fe/H]}^I_{N2}$, as already shown in the right panel of Fig.~5,
and so does not change the conclusions.

\begin{figure}[htb]
\vglue-2mm
\includegraphics[width=5.7cm]{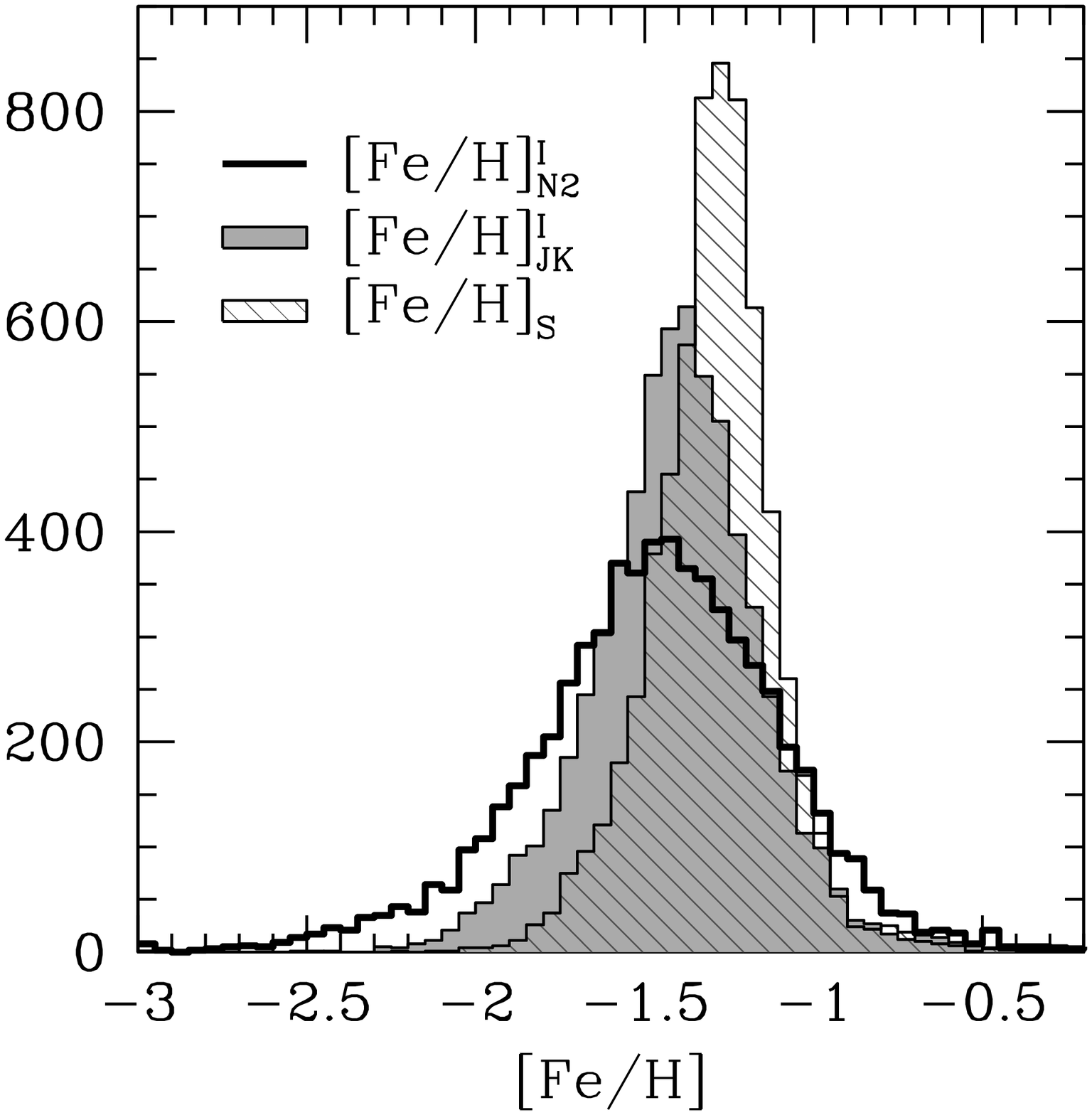}\hfil\includegraphics[width=5.7cm]{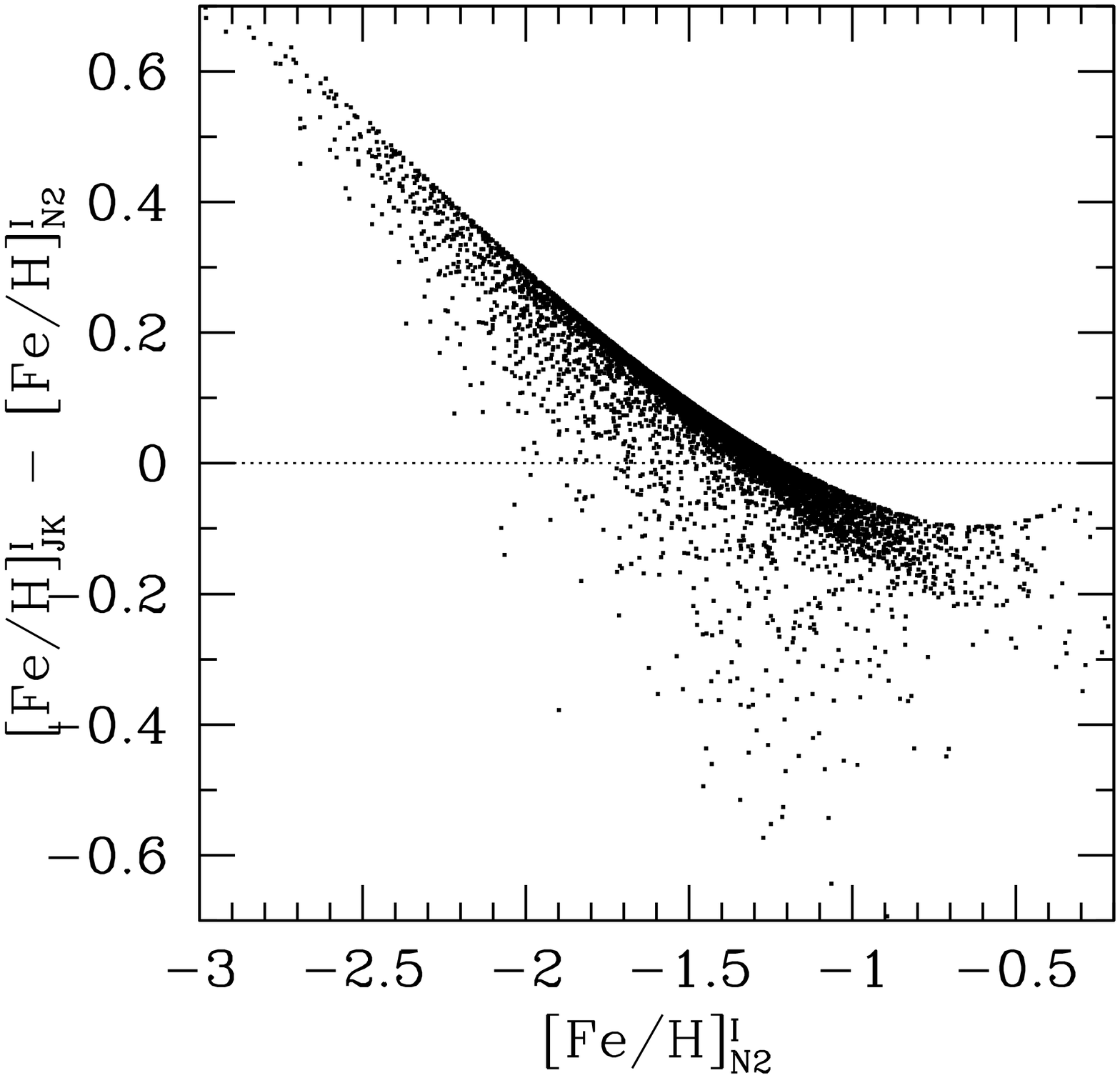}
\FigCap{{\it Left panel:} histograms of ${\rm [Fe/H]}^I_{N2}$,
${\rm [Fe/H]}^I_{\it JK}$, and ${\rm [Fe/H]}_S$ for the 6438 RRab. {\it Right panel:}
metallicity difference ${\rm [Fe/H]}^I_{\it JK}-{\rm [Fe/H]}^I_{N2}$ \vs
${\rm [Fe/H]}^I_{N2}$.}
\end{figure}

Deb \etal (2015) in their analysis of the SMC RRab sample from OGLE-III
decided to use the relation of JK96 to calculate iron abundances, rather
than that of N13, because they had found that formal errors on ${\rm
[Fe/H]}_{N}$ (Eq.~5a) are much larger than those on ${\rm
[Fe/H]}_{\it JK}$. However, this means overestimating [Fe/H] at low
metallicities. We think that larger errors on individual [Fe/H] values are
not as problematic when doing a statistical analysis of a large sample, as
an overestimation of the sample's global properties such as median iron
abundances and their gradients. Thus we decide to use the relation of
N13. The formal errors on ${\rm [Fe/H]}_N$ (Eq.~5a) are indeed
nonphysically large, so we will do not list them in this paper.  Instead, we
calculate errors for each star individually, as described in Section~6.

\vspace*{-7pt}
\Section{Photometric Metallicity in the Magellanic System}
\vspace*{-3pt}
We use the full dataset of 32 581 RRab stars from Soszyñski \etal (2016) to
create a photometric metallicity map of the Magellanic System. First, we
clean the sample by excluding objects with no {\it V}-band magnitude and
with large errors of phase parameters used for [Fe/H] estimation. We also
remove Blazhko stars, which we identify by the scatter of the light curve
around the Fourier model, and remove about 20\% of stars with the largest
scatter. This leaves 24\,133 objects in the final sample, among which
20\,573 belong to the LMC and 3560 to the SMC.

\subsection{Metallicity Distribution}
We then calculate photometric metallicities using the most accurate to date
relation of N13 (Eq.~5) after transforming it to the {\it I}-band with
Eq.(6). We decided to use a single relation for the ($\varphi^I_{31}
\rightarrow\varphi^V_{31}$) transformation rather than a set of equations
from Table~2, because the {\it V}-band data for many RRab are poor (see
discussion in Section~4.3). The formal errors of the N13 relation, using
the standard error propagation formula, become unrealistically large,
resulting in extremely high values of $\sigma_{\rm [Fe/H]}$. In such case
we decided to perform an error simulation. For each star, we randomly
select all coefficients in each equation, from their own distributions
(assuming they are Gaussian), and calculate ${\rm [Fe/H]}_i$. This is
repeated 1000 times for each star. The standard deviation of the resulting
distribution is then treated as the error of [Fe/H].

The distribution of metallicities is presented in Fig.~8 for both the LMC
(left panel) and the SMC (right panel). Median metallicity values are
${\rm [Fe/H]}_{J95}=-1.39\pm0.44$~dex on the HDS scale of J95,
and ${\rm[Fe/H]}_{ZW84}=-1.59\pm0.31$~dex on the Zinn and
West (1984, hereafter ZW84) scale. In the case of the SMC, median iron
abundances are ${\rm[Fe/H]}_{J95}=-1.77\pm0.48$~dex and ${\rm
  [Fe/H]}_{ZW84}=-1.85\pm0.33$~dex.  We used the equation of J95 in the
form: ${\rm [Fe/H]}_{ZW84}=({\rm [Fe/H]}_{J95}-0.88)/1.431$ to transform
from the HDS scale of J95 to the ZW84 scale.  Some studies use the relation
of Papadakis \etal (2000) for this purpose (${\rm [Fe/H]}_{ZW84}=1.028 {\rm
  [Fe/H]}_{J95}-0.242$), which gives similar results as the relation of J95
around ${\rm [Fe/H]}\simeq-1.4$~dex, but quickly offsets, \eg at ${\rm
  [Fe/H]}\simeq-2.0$~dex the difference between the two transformations is
$-0.23$~dex and at ${\rm [Fe/H]}\simeq0.0$~dex it is 0.53~dex. Papadakis
\etal (2000) do not explain in detail how their relation was obtained, so
it is not clear how reliable the relation is and what systematic errors it
may be subject to. For this reason we use the relation of J95 throughout
the paper.
\begin{figure}[htb]
\centerline{\includegraphics[width=8cm]{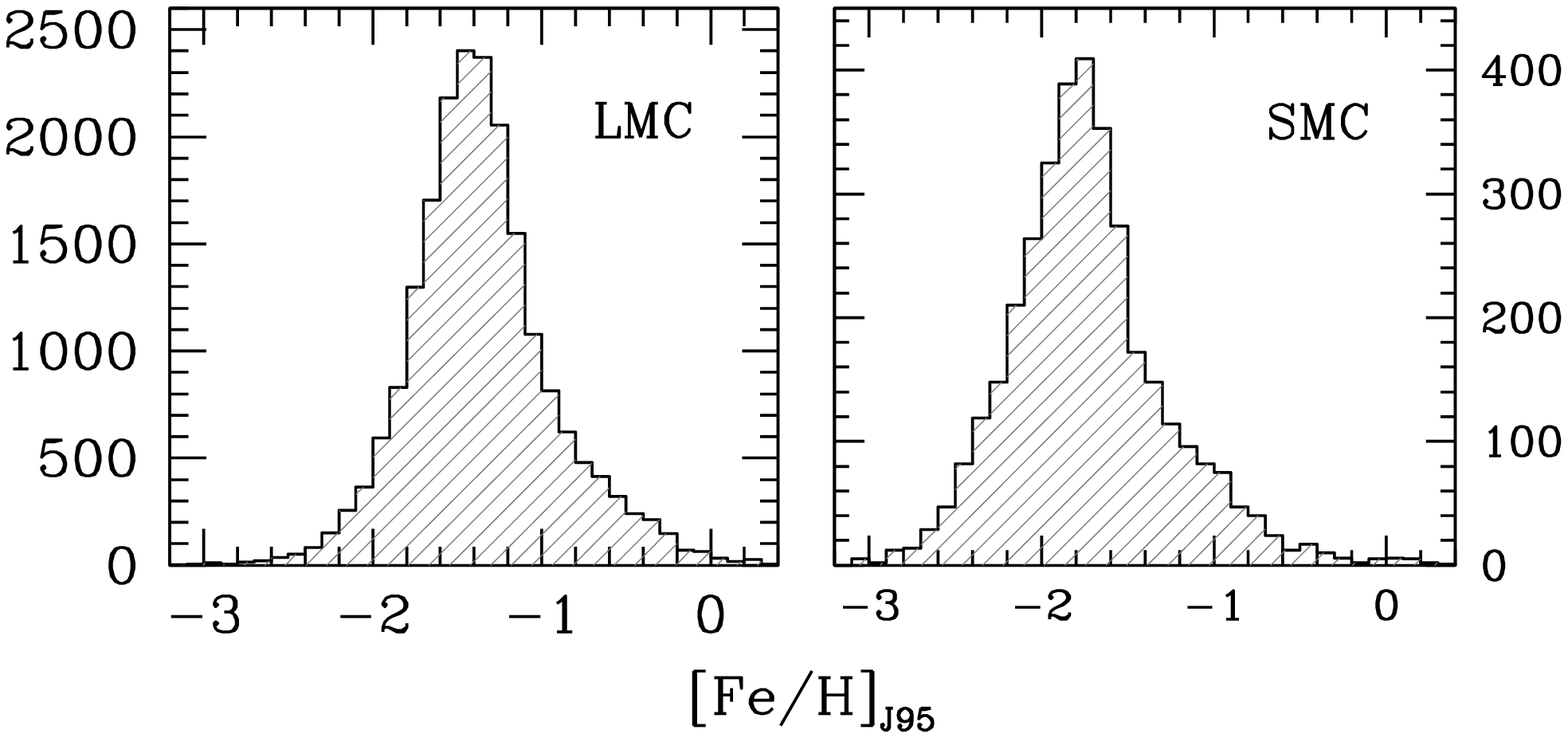}}
\FigCap{Histograms of ${\rm [Fe/H]}^I_{N}$ on the J95 metallicity scale.}
\end{figure}

Distances are calculated from the reddening-free Wesenheit index in
order to avoid correcting for extinction. The observed Wesenheit magnitude is: 
$$W=I-1.55\times(V-I)$$
and the absolute Wesenheit magnitude is:
$$W_0=-1.039-2.524\log P+0.147({\rm [Fe/H]}^I_{N,C}+0.04)$$
from Table~5 of Braga \etal (2015).  ${\rm [Fe/H]}^I_{N,C}$ is the iron
abundance on the Carretta \etal (2009) metallicity scale, which was
obtained by transforming ${\rm [Fe/H]}^I_N$ from the HDS scale of J95 to
the Carretta scale with: ${\rm [Fe/H]}_{C09}=1.001{\rm [Fe/H]}_{J95}-0.112$
(Kapakos \etal 2011). Median LMC distance is $D_{\rm LMC}=50.56$~kpc,
while median SMC distance is $D_{\rm SMC}=60.45$~kpc.  We also calculate
the three-dimensional coordinates for each object using transformation
equations from van der Marel and Cioni (2001) and Weinberg and Nikolaev
(2001):
\setcounter{equation}{7}
\begin{eqnarray}
        x &=&-d\times \cos(\delta)\sin(\alpha-\alpha_{\rm cen}) \\
        y &=& d\times(\sin(\delta)\cos(\delta_{\rm cen})-\cos(\delta)\sin(\delta_{\rm cen})\cos(\alpha-\alpha_{\rm cen})) \\
        z &=& d\times(\cos(\delta)\cos(\delta_{\rm cen})\cos(\alpha-\alpha_{\rm cen})+\sin(\delta)\sin(\delta_{\rm cen})) - D_{\rm cen}
\end{eqnarray}
where $d$ is the objects distance from observer, $\alpha_{\rm cen}$ and
$\delta_{\rm cen}$ are the LMC/SMC center equatorial coordinates and
$D_{\rm cen}$ is the median LMC/SMC distance. We adopt $\alpha_{\rm LMC} =
80.38$~deg, $\delta_{\rm LMC}=-69.61$~deg, and $\alpha_{\rm SMC}=
13.95$~deg, $\delta_{\rm SMC}=-72.78$~deg, which are the centers of the
distributions of $\alpha$ and $\delta$ in both galaxies. Having ($x,y,z$)
for each RRab we then calculate their distances from the LMC/SMC center as
$r=\sqrt{x^2+y^2+z^2}$. The three-dimensional distribution of OGLE-IV
RR~Lyr will be discussed in detail in Jacyszyn-Dobrzeniecka \etal (in
preparation).

Finally, in Fig.~9 we plot two-dimensional metallicity maps of the
Magellanic System. The top map shows all individual [Fe/H] values, with the
point size getting smaller toward galaxy centers, to increase the visible
resolution. In the bottom map, we plot averaged iron abundances in regions
where the number of stars in an area bin ($\approx0.335~{\rm deg^2}$) was at
least five, and leave individual measurements otherwise. By comparing two
maps we see that there is a large spread in [Fe/H] in every direction in
the Clouds, but after averaging, all areas seem to have a nearly constant
median [Fe/H] value. While the SMC does not seem to have a metallicity
gradient, we do see that the LMC has a higher average metallicity in the
center as compared to the outer regions.
\begin{figure}[htb]
\centerline{\includegraphics[width=12cm]{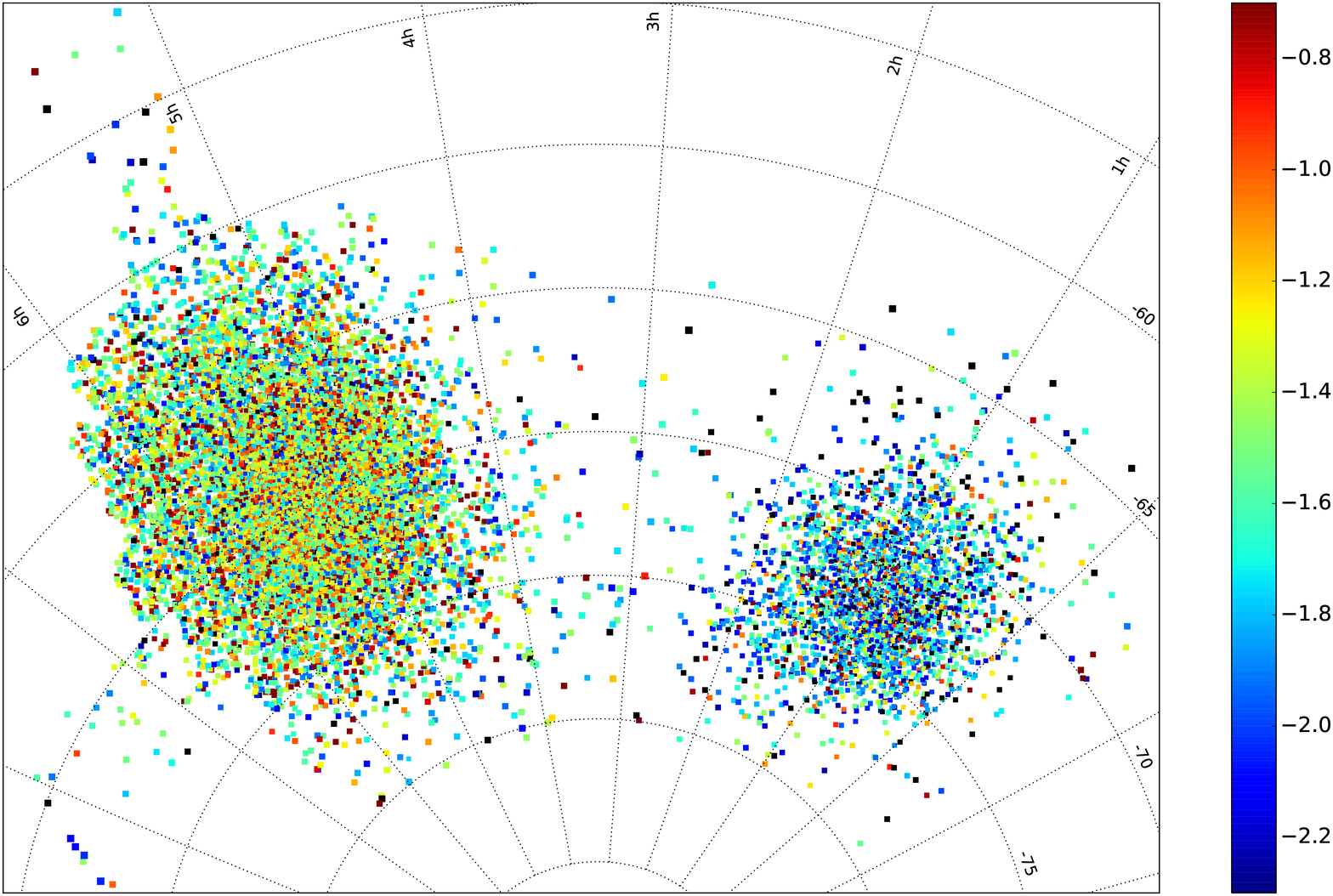}}
\centerline{\includegraphics[width=12cm]{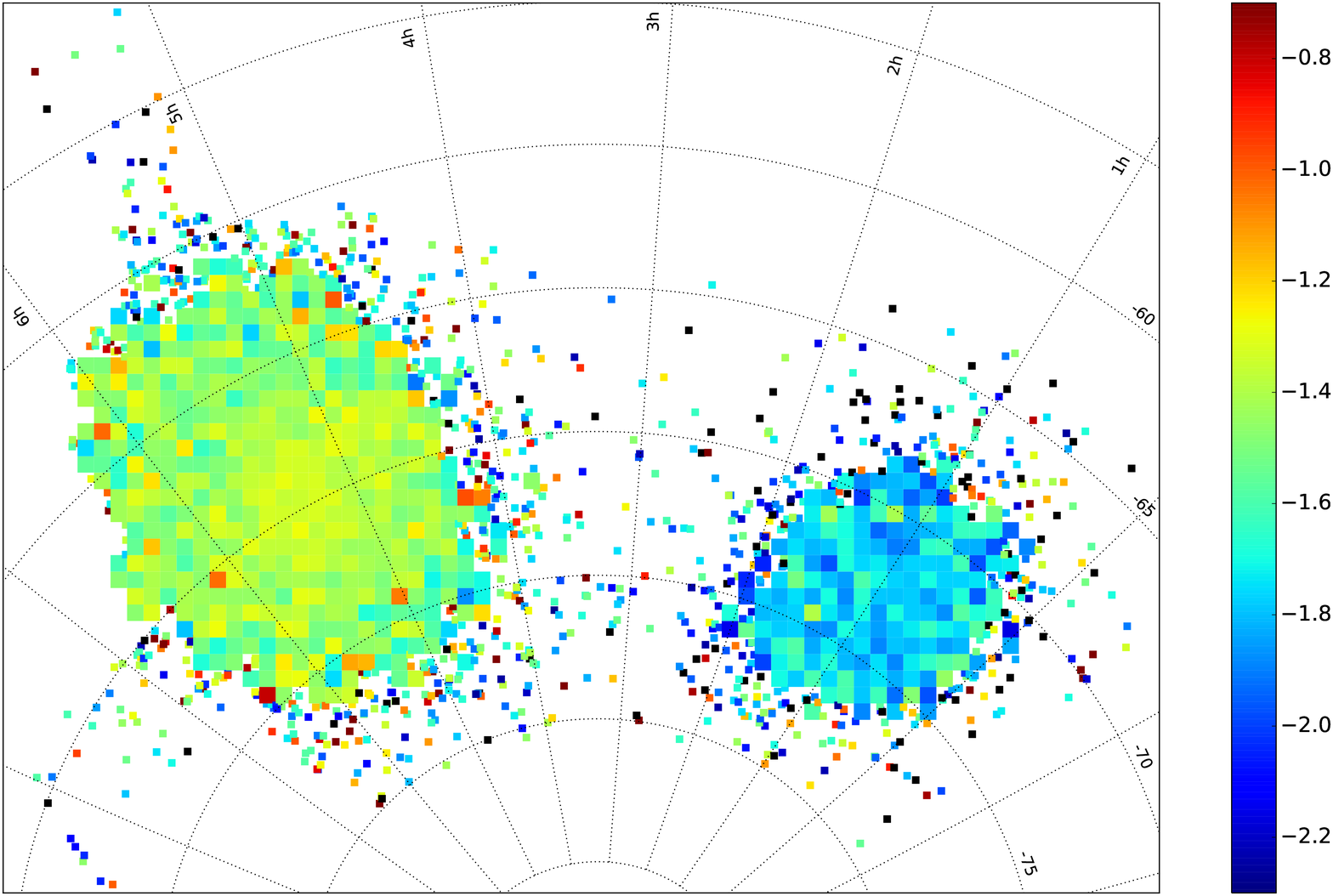}}
\FigCap{Metallicity maps of the Magellanic System on the J95 metallicity
  scale. {\it Top:} all individual [Fe/H] measurements, {\it bottom:} binned [Fe/H] in
  the galaxy centers and individual in the outskirts, where number of stars
  per area bin ($\approx0.335~{\rm deg^2}$) was less than five.}
\end{figure}

Metallicities and locations of all 24 133 RRab in our sample are available in
an electronic form from the OGLE website:

\centerline{\it http://ogle.astrouw.edu.pl}

The first few lines of the file are presented in Table~3.

\renewcommand{\arraystretch}{1.15}
\MakeTableee{|l|c|c|c|c|c|}{12.5cm}{Metallicities of 24 133 OGLE-IV RRab stars}
{\hline
\multicolumn{1}{|c|}{ID} & $\alpha$ [deg] & $\delta$ [deg] & D [kpc] & \multicolumn{1}{|c|}{${\rm [Fe/H]}_{J95}$} & \multicolumn{1}{|c|}{${\rm [Fe/H]}_{ZW84}$}  \\
\hline
lmc502\_08\_6887   &  81.655250 & -70.702472 & 51.65 & $-1.510 \pm 0.087$ & $-1.670 \pm 0.061$ \\
lmc563\_28\_7986   &  88.958750 & -65.657083 & 46.75 & $-1.470 \pm 0.039$ & $-1.642 \pm 0.027$ \\
lmc556\_21\_903    &  85.632917 & -65.586028 & 48.58 & $-0.601 \pm 0.451$ & $-1.035 \pm 0.315$ \\
lmc551\_14\_16904  &  86.225667 & -71.925306 & 49.26 & $-2.419 \pm 0.087$ & $-2.306 \pm 0.061$ \\
\multicolumn{1}{|c|}{...} & ... & ... & ... & ... & ...  \\
\hline
\noalign{\vskip3pt}
\multicolumn{6}{p{12cm}}{Full table is available for download from the
OGLE website: {\it http://ogle.astrouw.edu.pl}}
}

\subsection{Metallicity Gradient}
In order to better see whether there is a metallicity gradient in the
galaxies, in Fig.~10 we plot [Fe/H] \vs radial distance $r$ from the
LMC/SMC center. Gray dots show all 24\,133 RRab variables used in the
analysis, 20\,573 in the LMC (left) and 3560 in the SMC (right). Large gray
circles mark median metallicity values in 1~kpc wide bins, requiring at
least 40 points in the bin in the case of the LMC and 20 in the case of the
SMC. Dotted lines mark $1\sigma$ and $3\sigma$ deviations from the median
value in a given bin. While there appears to be no metallicity gradient in
the SMC, we clearly see that in the LMC [Fe/H] decreases with distance from
the center. The decline is evident within 4~kpc from the galaxy center,
then the metallicity remains constant until about 8~kpc, when it shows
another mild decline. The slope of the gradient is $-0.029\pm0.002$~dex/kpc
in the inner 5~kpc and $-0.030\pm0.003$~dex/kpc beyond 8~kpc. For the
entire galaxy, the slope is $-0.019\pm0.002$~dex/kpc.
\begin{figure}[htb]
\includegraphics[width=6cm]{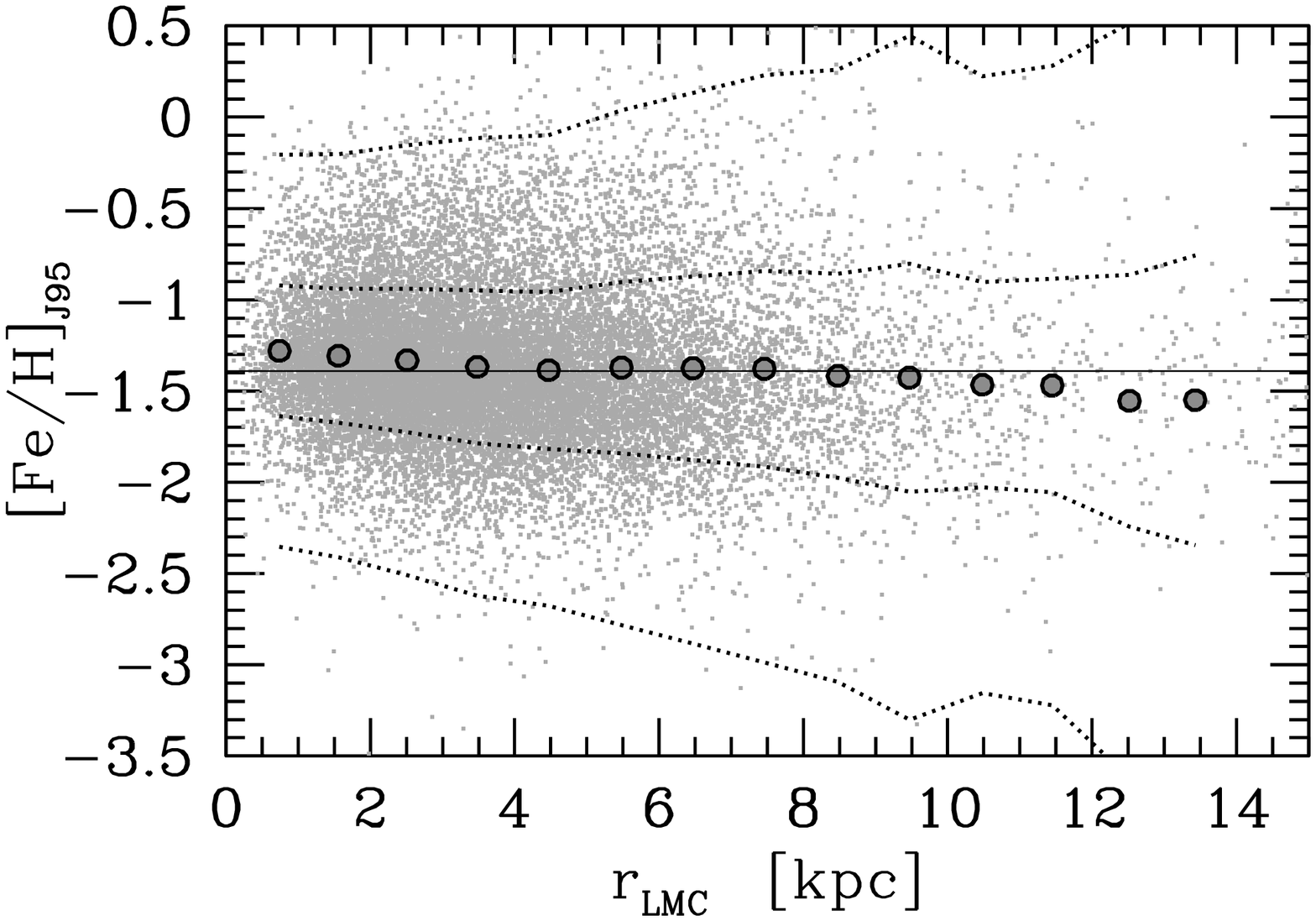}\hfil\includegraphics[width=6cm]{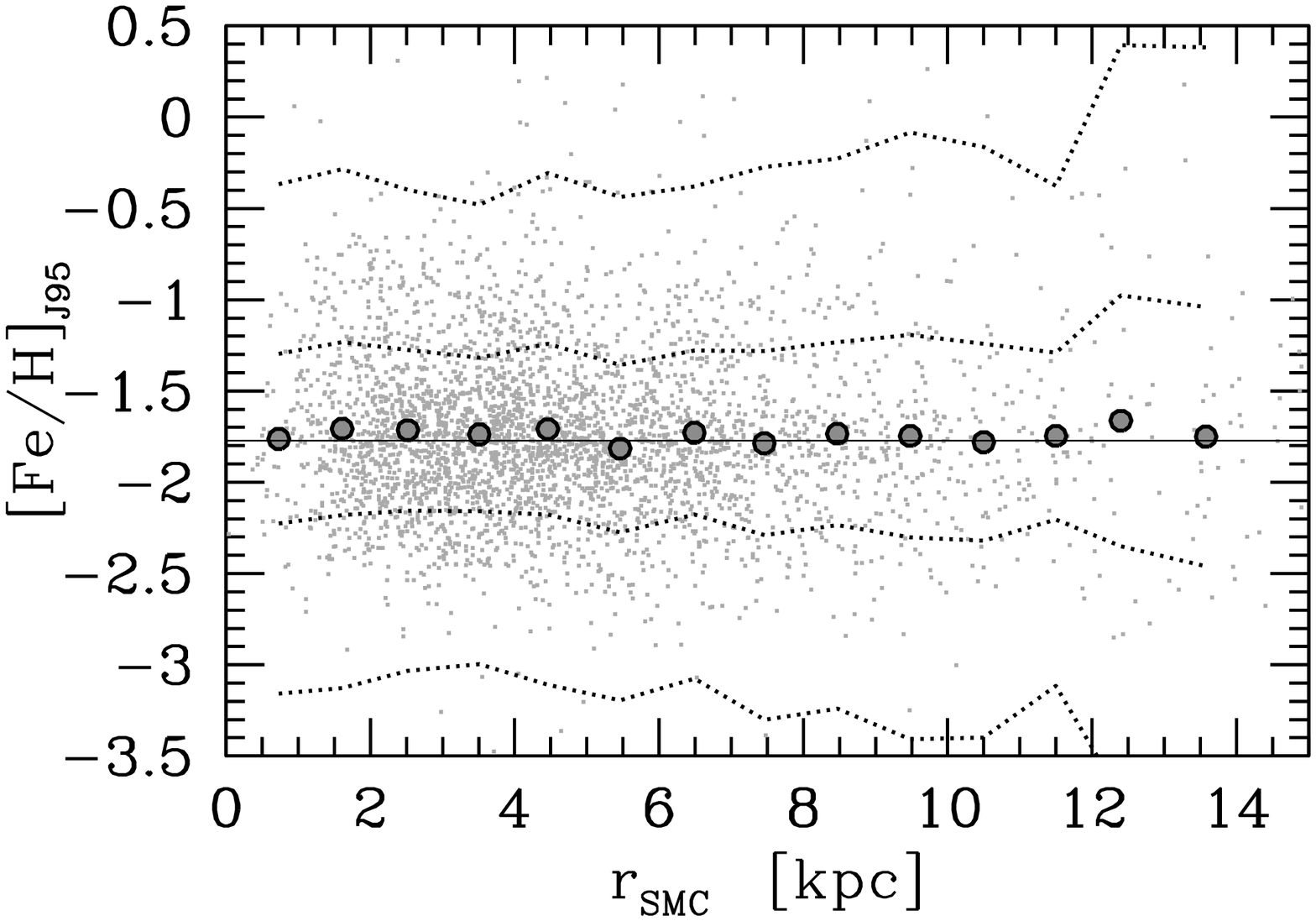}
\FigCap{Metallicity \vs distance from the galaxy center for the LMC ({\it
left}) and the SMC ({\it right}). Gray dots show all 24\,133 RRab used
in the analysis. Large gray circles mark median metallicity values in
1~kpc wide bins, requiring at least 40 points in the bin in the case of
the LMC and 20 in the case of the SMC. Solid line shows the median iron
abundance of $-1.39$~dex in the LMC and $-1.77$~dex in the SMC. Dotted
lines mark $1\sigma$ and $3\sigma$ deviations from the median value.}
\end{figure}

\begin{figure}[b]
\centerline{\includegraphics[width=6cm]{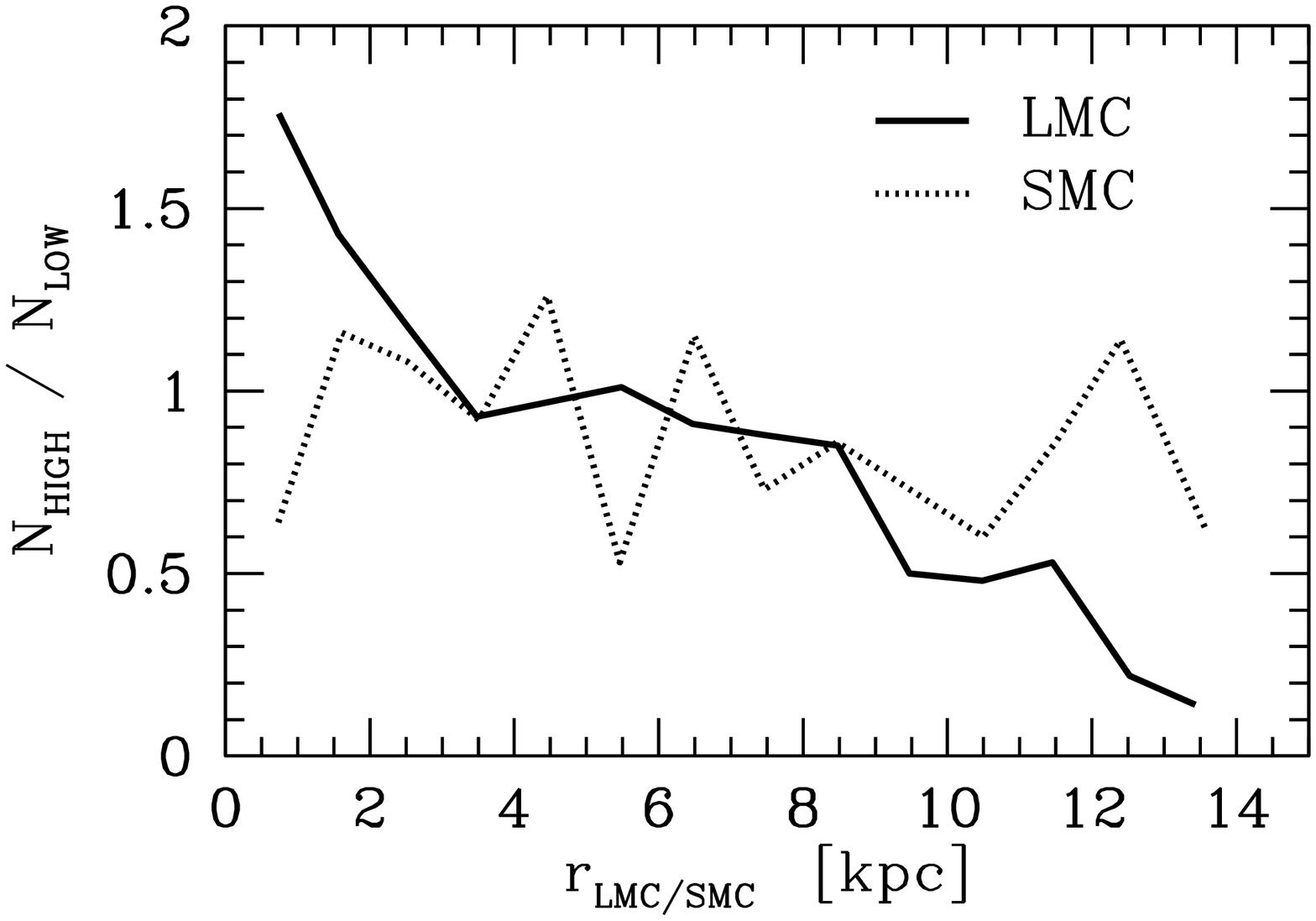}}
\FigCap{Ratio of a number of stars \vs distance from the galaxy center for
the LMC (solid line) and the SMC (dotted line). $N_{\rm HIGH}$ is the
number of stars for which ${\rm [Fe/H]}>F_1$ while $N_{\rm LOW}$ is the
number of stars for which ${\rm [Fe/H]}<F_2$. In the case of the LMC
$F_1=-0.55$~dex and $F_2=-2.0$~dex, while for the SMC $F_1=-1.3$~dex and
$F_2=-2.15$~dex.}
\end{figure}
\begin{figure}[p]
\centerline{\includegraphics[width=5.9cm]{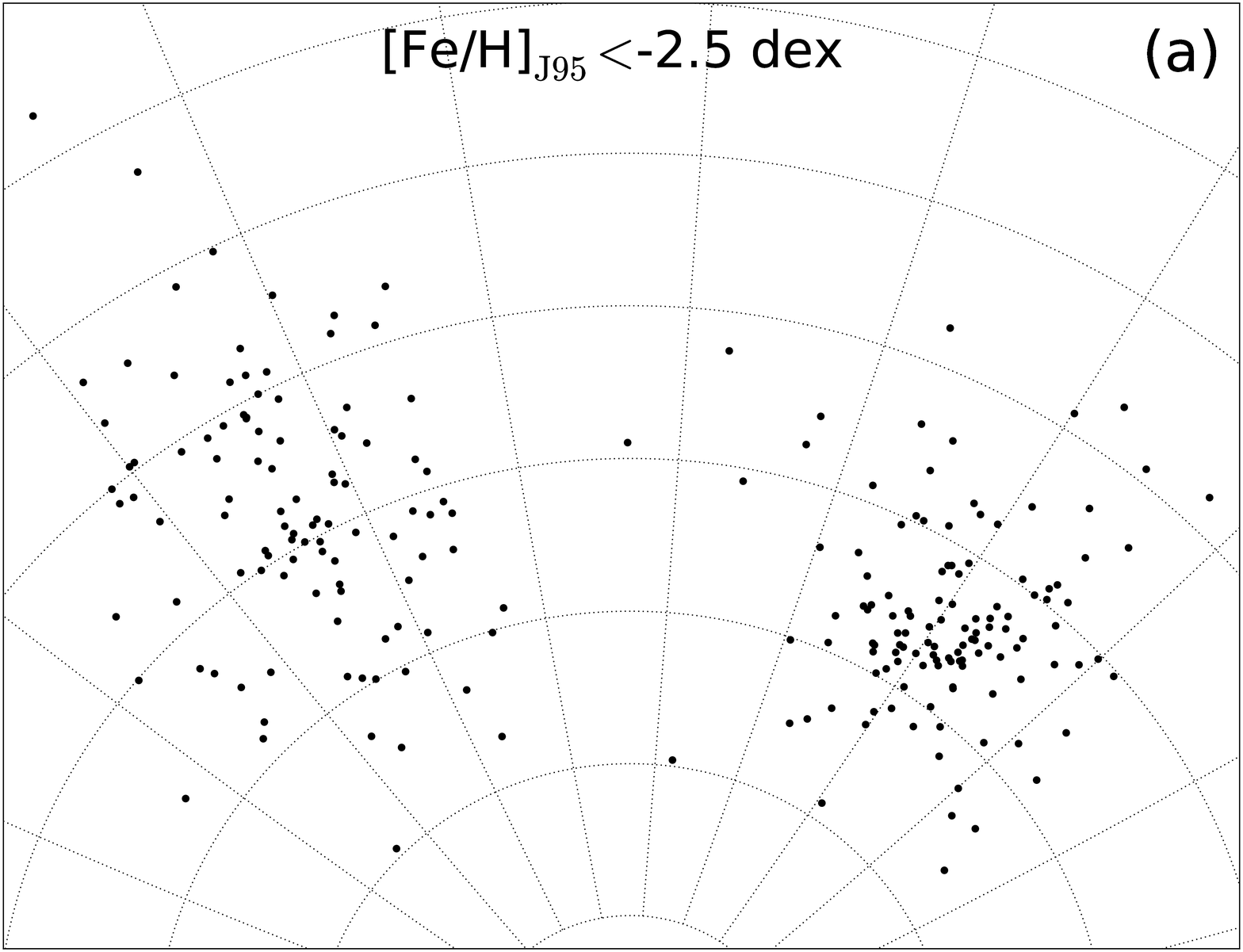}\hfil\includegraphics[width=5.9cm]{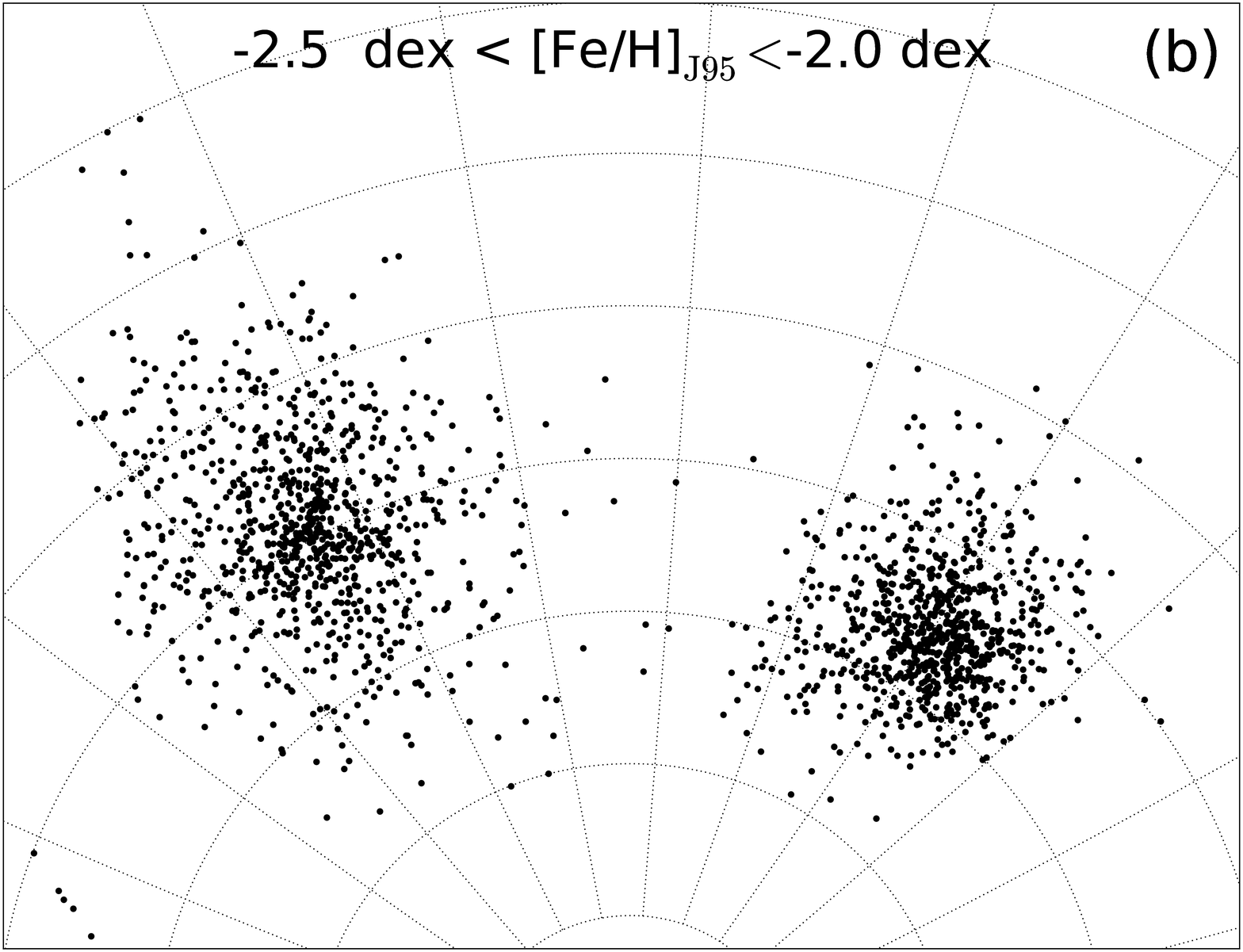}}
\vspace*{4mm}
\centerline{\includegraphics[width=5.9cm]{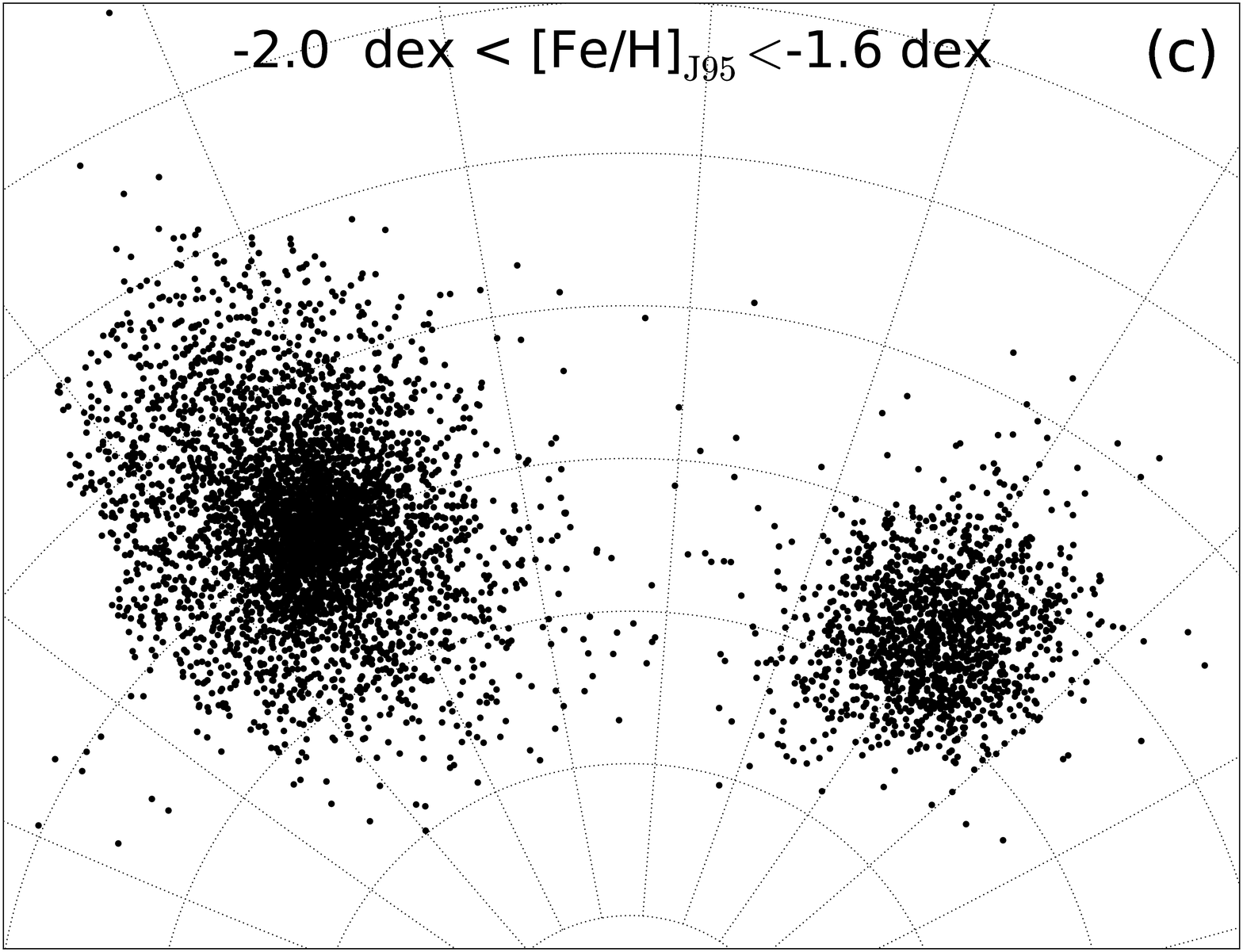}\hfil\includegraphics[width=5.9cm]{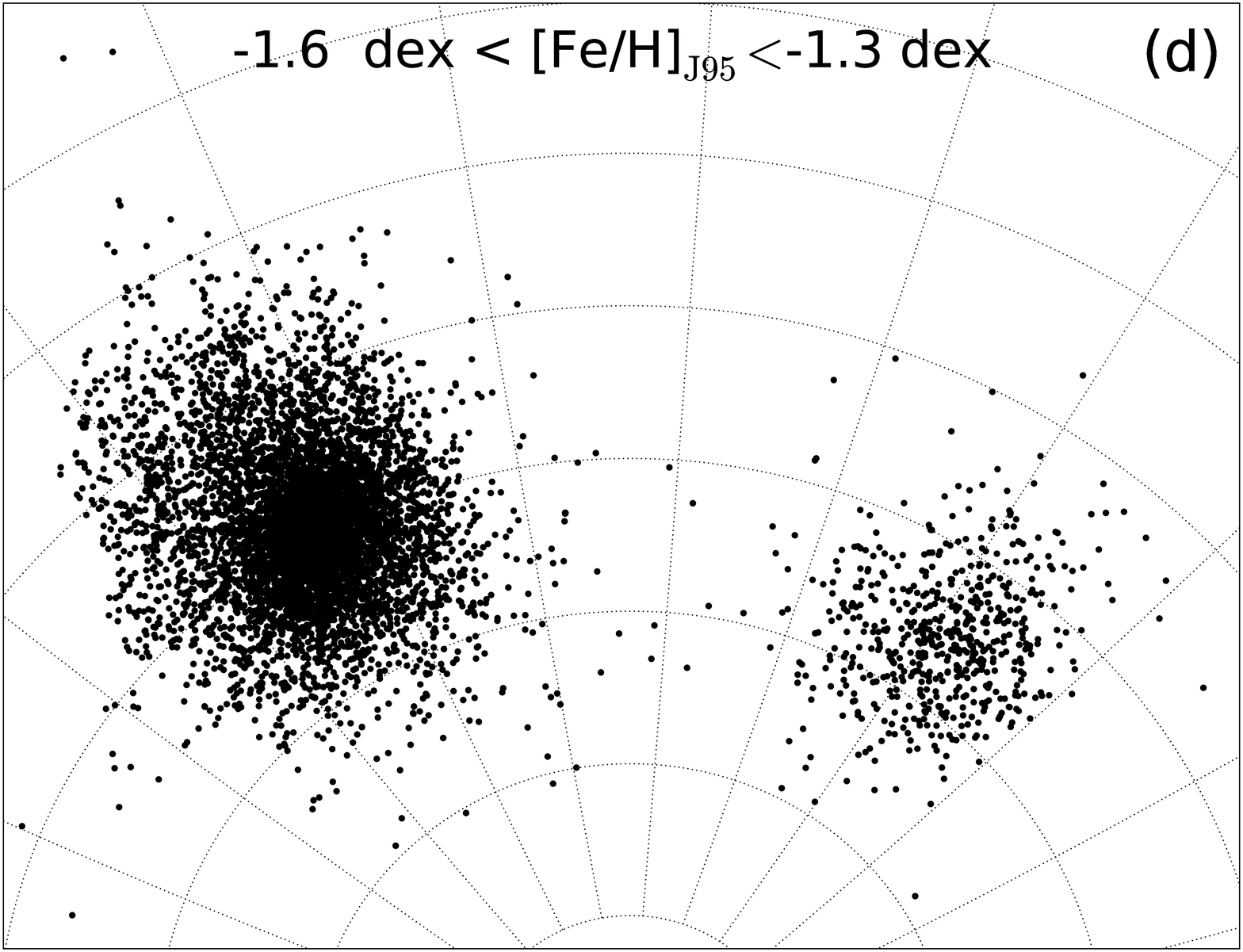}}
\vspace*{4mm}
\centerline{\includegraphics[width=5.9cm]{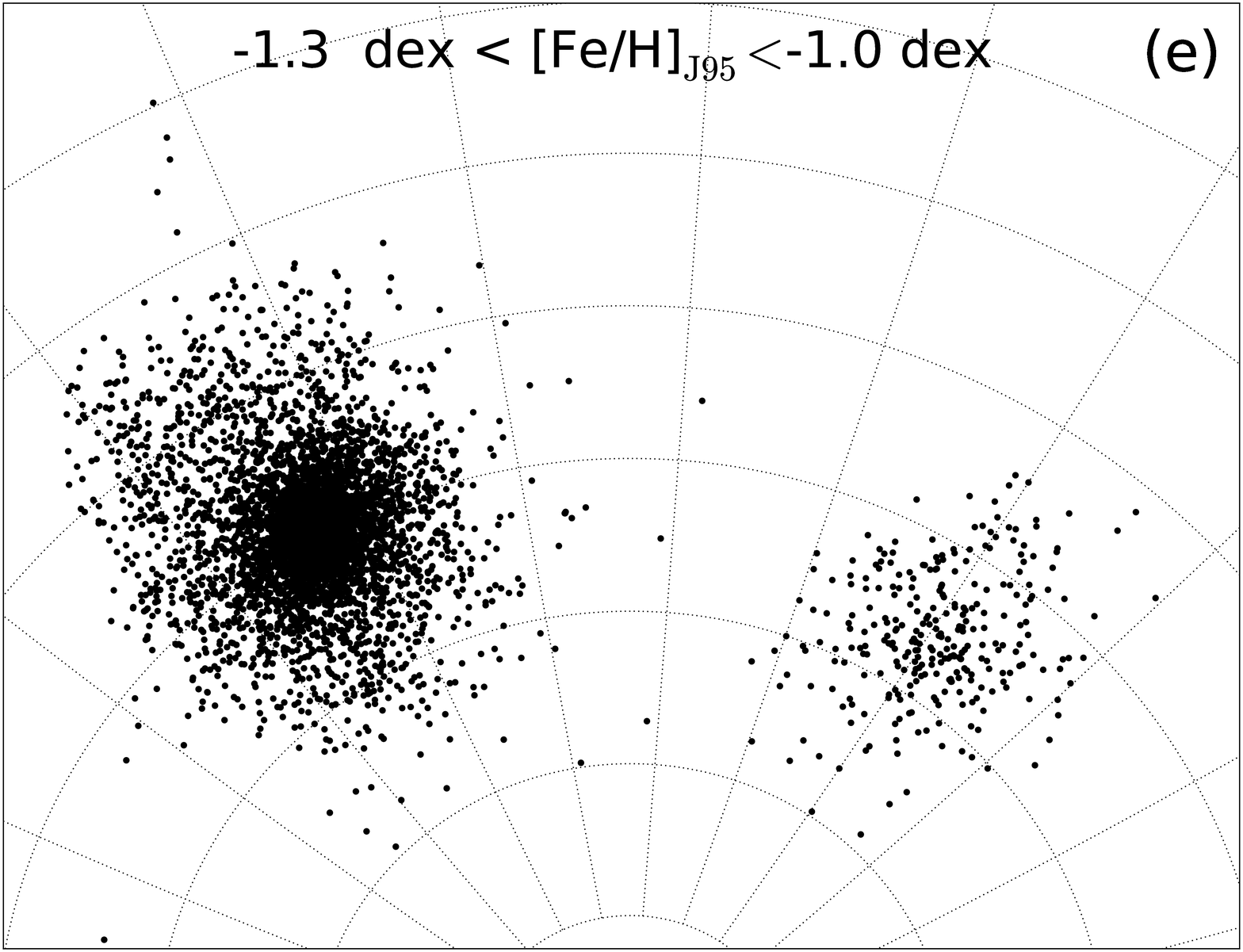}\hfil\includegraphics[width=5.9cm]{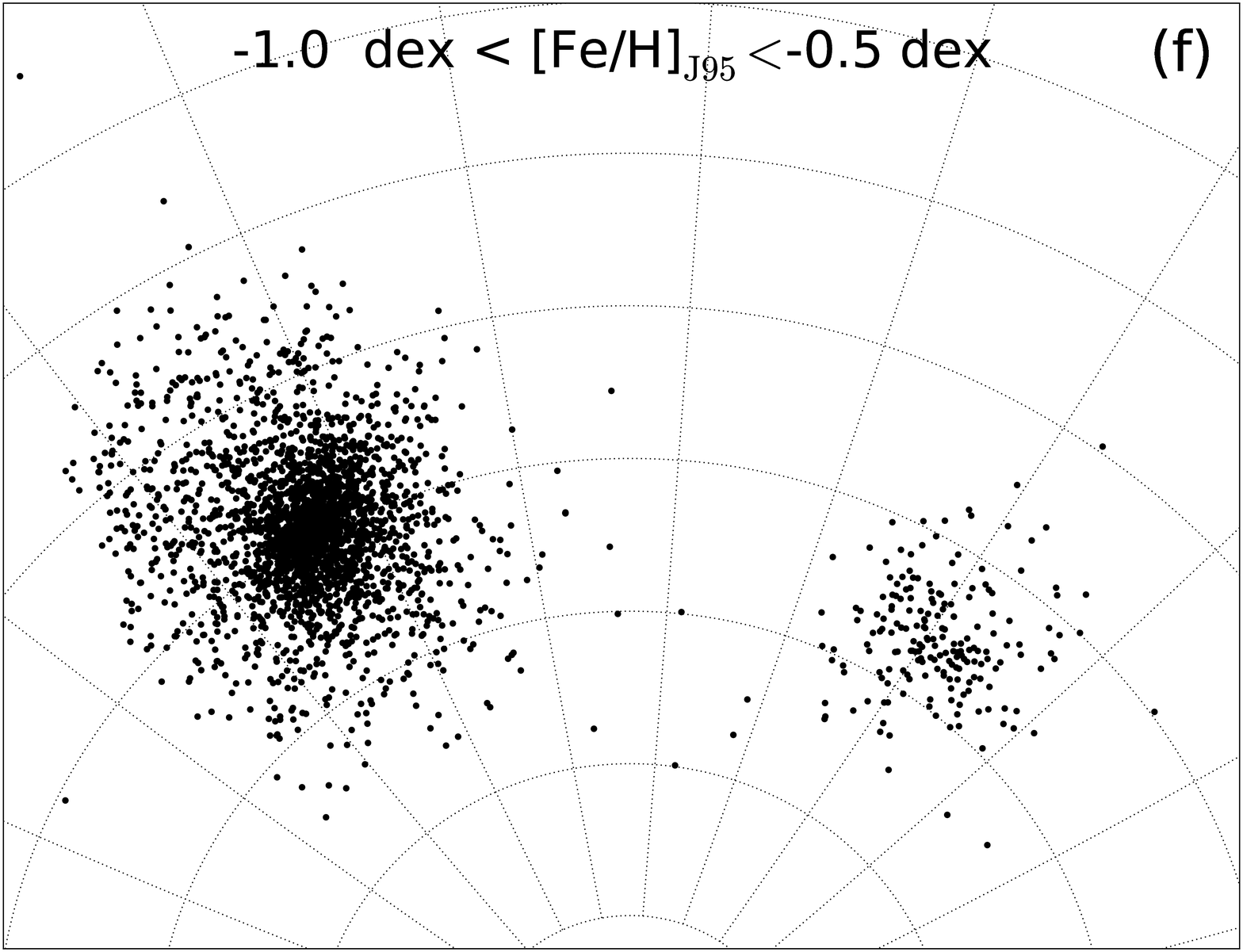}}
\vspace*{4mm}
\centerline{\includegraphics[width=5.9cm]{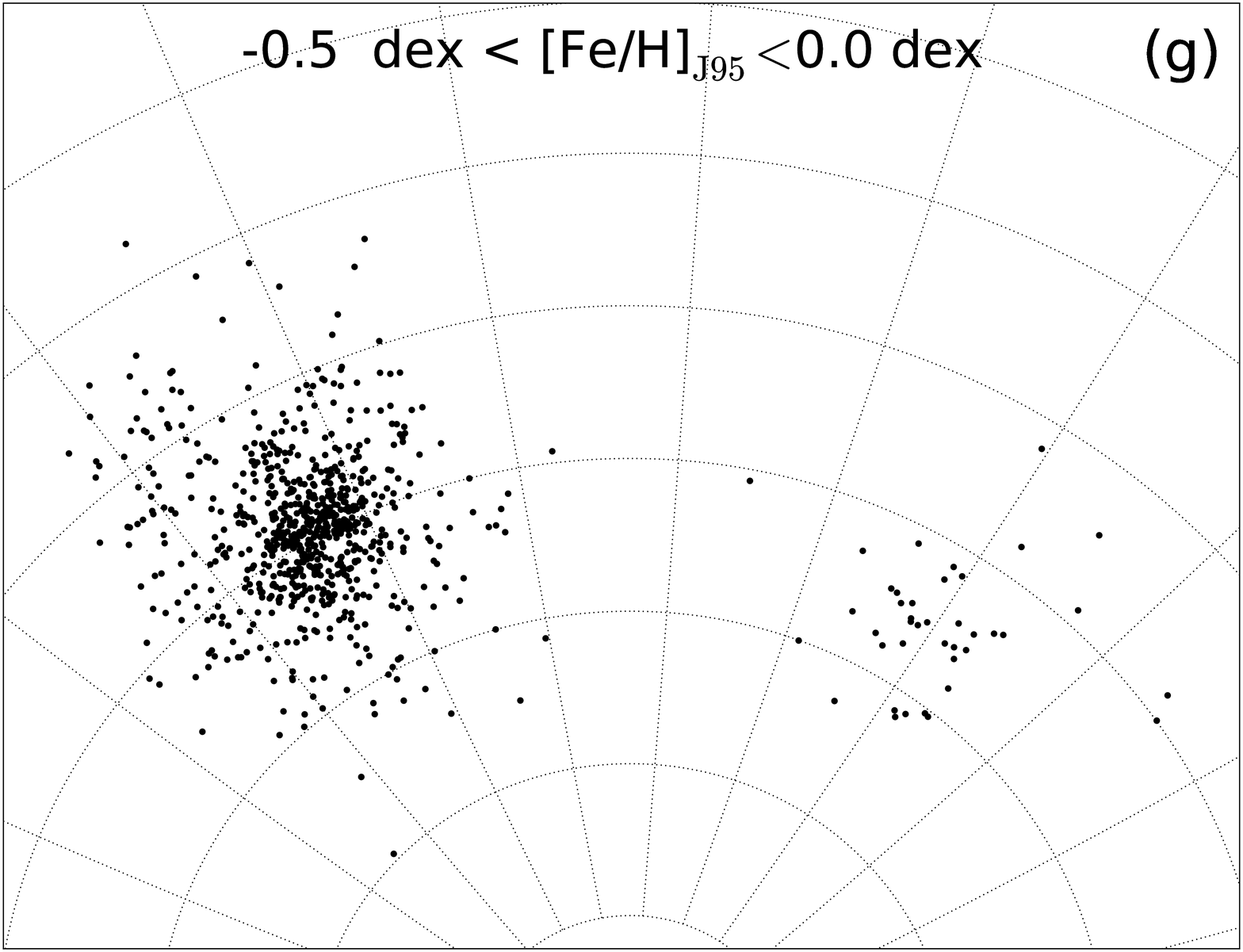}\hfil\includegraphics[width=5.9cm]{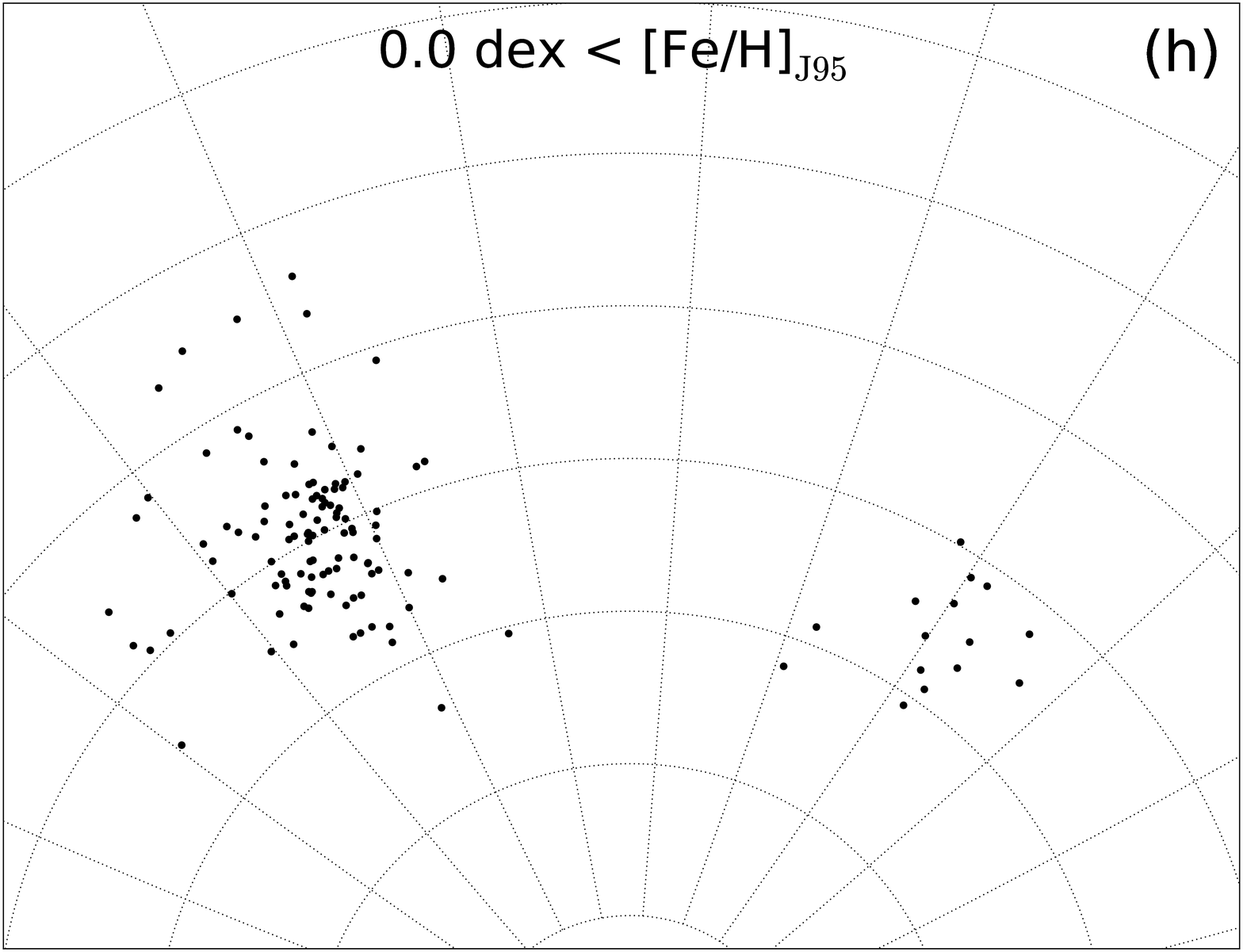}}
\vskip5pt
\FigCap{Metallicity tomography of the Magellanic System, starting from low
metallicities ({\it panel~a}) to high metallicities ({\it panel~h}).}
\end{figure}
The significance of the result could be argued due to the large scatter
(and errors) in [Fe/H], but Fig.~11 further supports the finding. The plot
shows how the ratio of a number of stars with high metallicities $N_{\rm
HIGH}$ to a number of stars with low metallicities $N_{\rm LOW}$, changes
with the distance from the center of each galaxy $r_{\rm LMC/SMC}$. In
other words, $N_{\rm HIGH}$ is the number of stars for which ${\rm
[Fe/H]}>F_1$ while $N_{\rm LOW}$ is the number of stars for which ${\rm
[Fe/H]}<F_2$ in a given distance bin. In the case of the LMC
$F_1=-0.55$~dex and $F_2=-2.0$~dex, while for the SMC $F_1=-1.3$~dex and
$F_2=-2.15$~dex. $F_1$ and $F_2$ were chosen such that the ratio $N_{\rm
HIGH}/N_{\rm LOW}=1$ for the entire galaxy and that the number of stars
$N_{\rm HIGH}$ and $N_{\rm LOW}$ in each distance bin was reasonable.
Fig.~11 shows that the center of the LMC has more metal-rich RRab than the
outskirts of this galaxy, suggesting an existence of the metallicity
gradient.  As noted before, [Fe/H] declines within $\approx4~$kpc from the
center, then plateaus until about 8~kpc when the second decline begins.  On
the other hand, no significant metallicity change is observed in the SMC,
as concluded from Fig.~10.

Last, in Fig.~12 we show the two dimensional distribution of RRab stars
from eight metallicity bins, starting from low metallicity ${\rm [Fe/H]}
<-2.5$~dex, panel~(a), and ending at high metallicity ${\rm [Fe/H]}
>0$~dex, panel~(h). Note that the metallicity range within a bin varies
from 0.5~dex to 0.3~dex, in order to compensate for higher stellar density
around [Fe/H] distribution maxima. In the case of the LMC, metal-rich RRab
stars prefer the centers of the galaxies, while the metal-poor are evenly
distributed. In the case of the SMC, this is not as well pronounced, but we
do see that the most metal-rich RRab (panel~h) are more concentrated than
most metal-poor RRab (panel~a). The Magellanic Bridge RRab pulsators tend
to have lower metallicities (panels a to d), with a slighter preference of
lower metallicities typical for the SMC ($-2.0<{\rm [Fe/H]<-1.6}$~dex,
panel~(c)). We do not observe high metallicity RRab stars in the Magellanic
Bridge (panels g and h).

\subsection{Is the Metallicity Gradient Real?}
Photometric metallicities in the Magellanic System, presented in this
section, were calculated using the method of N13 and the metallicity
independent transformation between $\varphi^I_{31}$ and
$\varphi^V_{31}$. As discussed in Section~4.3, choosing a metallicity
independent over the metallicity dependent ($\varphi^I_{31}\rightarrow
\varphi^V_{31}$) interrelation, results in overestimating [Fe/H] at the low
end of the distribution and underestimating it at the high end. While the
mean metallicity values for the LMC and the SMC are not affected, the
metallicity gradient may be sensitive to such systematic changes. This
poses the question about the reality and validity of our results.

In order to check how the choice of the photometric metallicity calculation
method affects the metallicity distribution, we repeat calculations from
Sections~5.1 and~5.2, but this time using the relation of S05 (Eq.~3),
which was calibrated on the {\it I}-band light curves and thus does not
need any additional transformations. Median and standard deviation values
of the resulting distributions are $\langle{\rm
[Fe/H]}_S\rangle=-1.27\pm0.25$~dex for the LMC and $\langle{\rm
[Fe/H]}_S\rangle=-1.48\pm0.26$~dex for the SMC. As already shown in
the left panel of Fig.~7 and discussed in Section~4.4, the S05 method
produces the highest mean [Fe/H] of all methods, with the lowest
dispersion, and significantly overestimates [Fe/H] at the low metallicity
end, as compared to ${\rm [Fe/H]}_{\it JK}$ and ${\rm [Fe/H]}_N$.

The ${\rm [Fe/H]}_S$ metallicity distribution in the Magellanic Clouds is
very similar to that presented in Fig.~9, only its high and low end limits
are different, so we do not include an additional figure to illustrate
this. What is interesting, is whether there is a change in [Fe/H] with the
distance from the centers of the two galaxies. In Fig.~13 we show the
metallicity gradient both in the LMC and the SMC. The general conclusion is
very similar to that from Fig.~11 -- we observe a metallicity gradient in
the LMC, while there seems to be no gradient in the SMC. The difference is
that Fig.~13 does not have the plateau seen in Fig.~11, and the slope of
the gradient is lower: $-0.011\pm0.001$~dex/kpc as opposed to
$-0.019\pm0.002$~dex/kpc (total). Such differences are expected since the
two distributions are different. The presence of the metallicity gradient
in the LMC, independently of the method used, confirms its reality.
\begin{figure}[htb]
\centerline{\includegraphics[width=6cm]{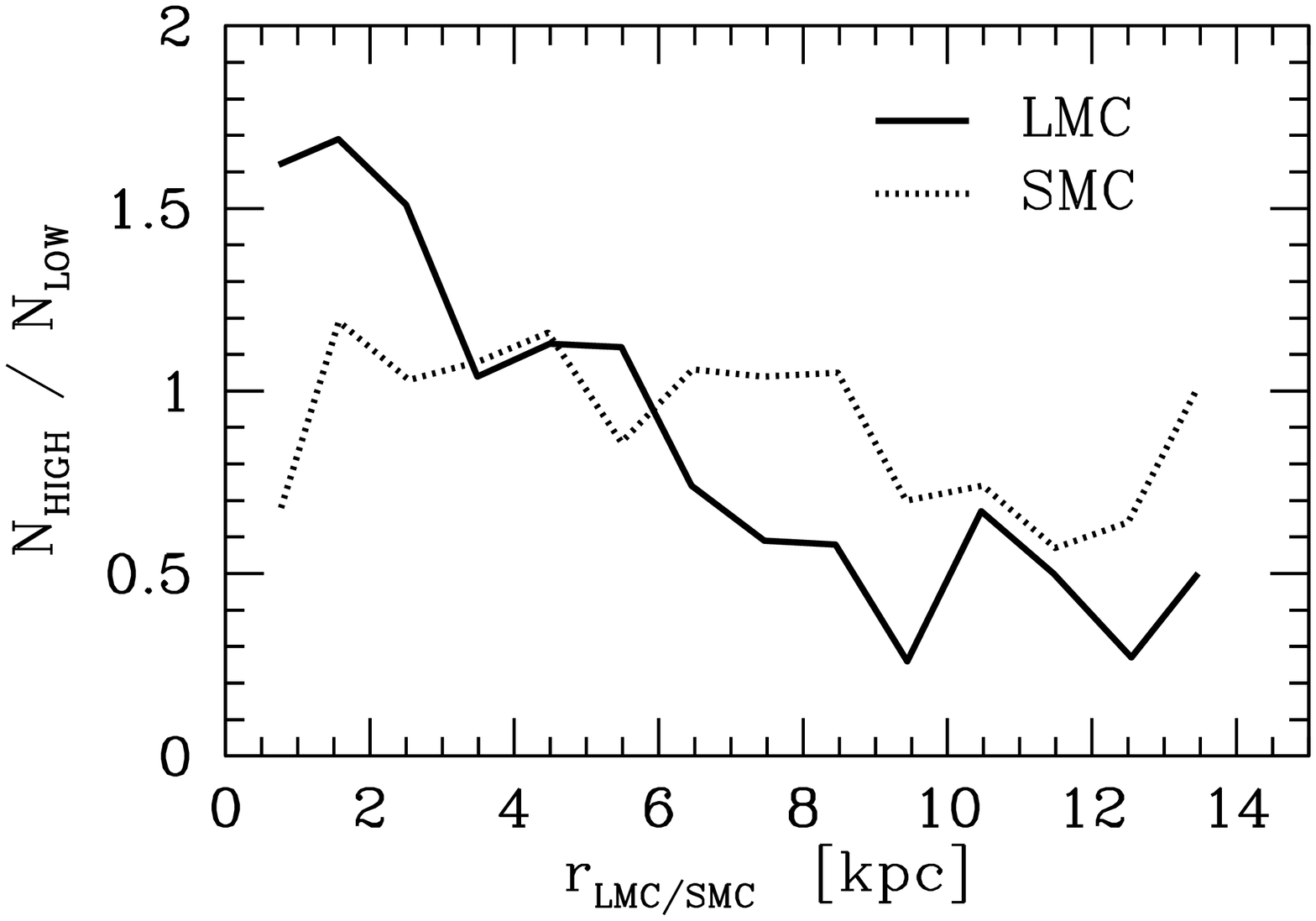}} 
\vskip3pt
\FigCap{Similar as Fig.~11, but for the ${\rm [Fe/H]}_S$ photometric
metallicity. Ratio of a number of stars \vs distance from the galaxy
center for the LMC (solid line) and the SMC (dotted line). $N_{\rm
HIGH}$ is the number of stars for which ${\rm [Fe/H]>F_1}$ while
$N_{\rm LOW}$ is the number of stars for which ${\rm [Fe/H]}<F_2$. In
the case of the LMC $F_1=-0.8$~dex and $F_2=-1.65$~dex, while for the
SMC $F_1=-1.37$~dex and $F_2=-1.6$~dex.}
\end{figure}

\section{Comparison with Previous Photometric Metallicity Studies of the Magellanic Clouds}
\subsection{Metallicity Distribution}
There have been numerous studies of photometric metallicity in the
Magellanic Clouds using RR~Lyr type stars. Methods for [Fe/H] calculation
used by various authors were either those using period and amplitude (by
Alcock \etal 2000 and Sandage 2004) or those using a linear relation
between period and $\varphi_{31}$ (by JK96 and S05). Since this is the
first work that applies the relation of N13 to construct the photometric
metallicity map of the Magellanic System, we may expect to see lower on
average metallicities than in other publications, as the new relation
provides lower metallicity values at the low end of the [Fe/H]
distribution.

\renewcommand\arraystretch{1.15}
\SetTableFont{\scriptsize}
\MakeTableee{|l|l|l|c|c|c|c|}{12.5cm}{Photometric metallicity in the Magellanic Clouds from RRab stars}
{
\hline
\multirow{2}{*}{REFERENCE} & \multirow{2}{*}{${\rm EQN.}^{*}$} & \multirow{2}{*}{DATA} & \multicolumn{2}{|c|}{LMC} & \multicolumn{2}{|c|}{SMC}  \\
\cline{4-7}
 & &  & \multicolumn{1}{|c|}{J95} & \multicolumn{1}{|c|}{ZW84} & \multicolumn{1}{|c|}{J95} & \multicolumn{1}{|c|}{ZW84}  \\
\hline
Deb and Singh (2010) & JK96 &O-II &        --        &        --        & $-1.56 \pm 0.25$ &        --        \\
\hline
Kapakos \etal (2011) & JK96 &O-III&        --        &        --        & $-1.51 \pm 0.41$ &        --        \\
Kapakos \etal (2012) & JK96 &O-III&        --        &        --        & $-1.58 \pm 0.41$ &        --        \\
\hline
Haschke \etal (2012) & S05  &O-III& $-1.22 \pm 0.26$ & $-1.49 \pm 0.26$ & $-1.42 \pm 0.33$ & $-1.70 \pm 0.33$ \\
\hline
Wagner-Kaiser and    &A00&O-III& -- & $-1.70 \pm 0.25$ &        --        &      --        \\
Sarajedini (2013)                      &S05&O-III& -- & $-1.63 \pm 0.21$ &       --        &        --        \\
\hline
Deb and Singh (2014) & JK96 &O-III&        --        & $-1.57 \pm 0.12$ &        --        &        --        \\
                     & S05  &O-III&        --        & $-1.50 \pm 0.12$ &        --        &        --        \\
                     & A00  &O-III&        --        & $-1.67 \pm 0.18$ &        --        &        --        \\
                     & Sd04 &O-III&        --        & $-1.57 \pm 0.20$ &        --        &  --  \\
\hline
This work            & N13  &O-IV& $-1.39 \pm 0.44$ & $-1.59 \pm 0.31$ & $-1.77 \pm 0.48$ & $-1.85 \pm 0.33$ \\
\hline
\noalign{\vskip5pt}
\multicolumn{7}{p{11cm}}{$^*$Methods of metallicity calculation: A00 -- Alcock \etal (2000),
JK96 -- Jurcsik and Kovacs (1996), S05 -- Smolec (2005), Sd04 -- Sandage (2004), N13 -- Nemec \etal (2013)}}

\begin{figure}[htb]
\includegraphics[width=6cm]{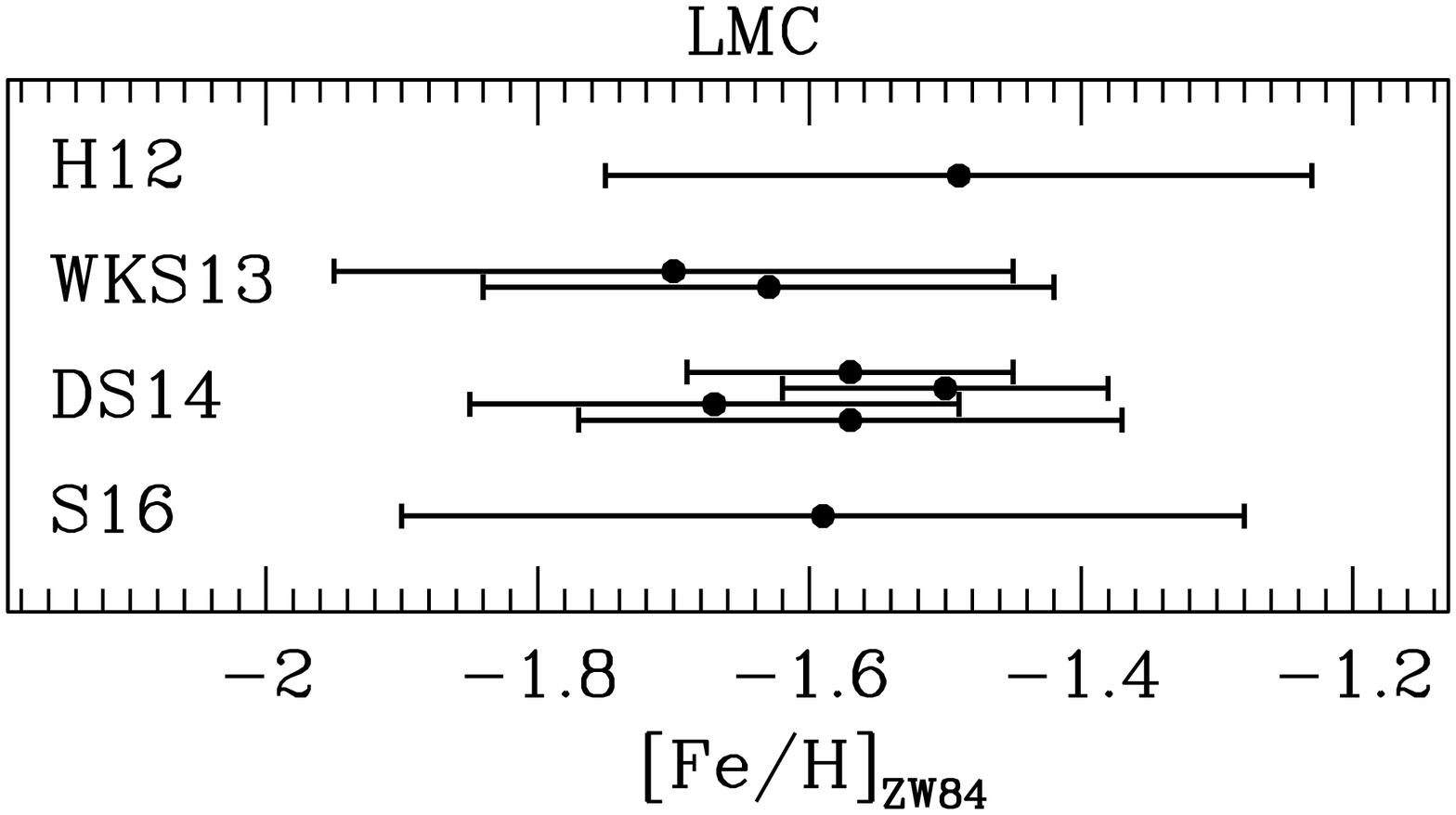}\hfil\includegraphics[width=6cm]{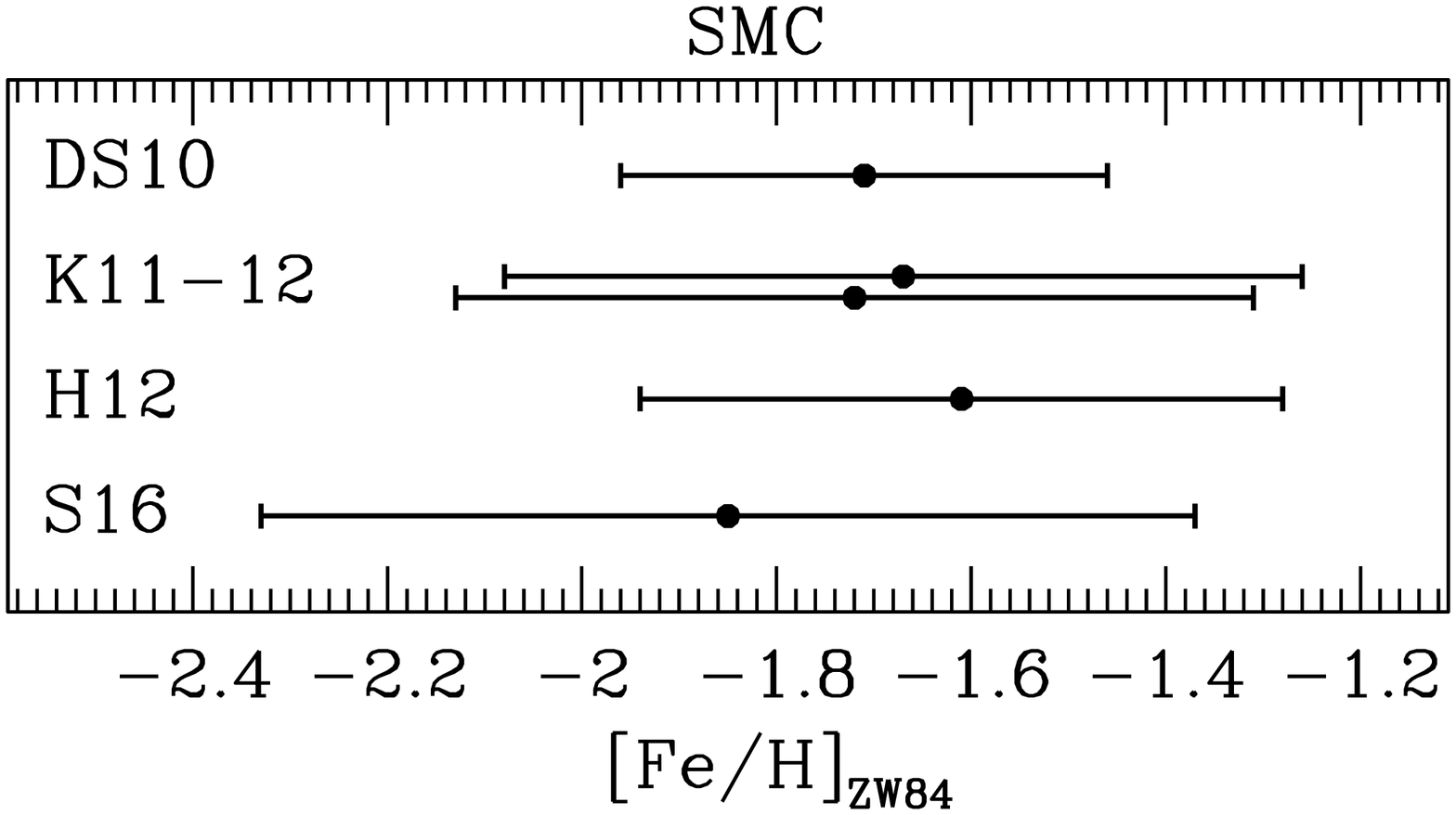}
\vskip5pt
\FigCap{Different [Fe/H] estimations for the LMC ({\it left}) and the SMC
({\it right}) from the literature and this work, as listed in
Table~4. Symbols are as follows: DS10~--~Deb and Singh (2010),
DS14~--~Deb and Singh (2014), H12~--~Haschke \etal (2012), K11~--~Kapakos
\etal (2011), K12~--~Kapakos and Hatzidimitriou (2012),
WKS13~--~Wagner-Kaiser and Sarajedini (2013), S16~--~this work.  Note
that all points are on the ZW84 metallicity scale -- SMC points from
Table~4 have been transformed from the J95 metallicity scale with the
same equation (by J95).}
\end{figure}

Mean metallicity values from the literature are listed in Table~4, in the
J95 and ZW84 metallicity scale, depending upon availability. In the case of
the LMC, all the cited studies provide [Fe/H] in the ZW84 scale and only
one in the scale of J95, while in the case of the SMC it is the
opposite. Table~4 also provides information about the method and data used
to calculate the iron abundances. Note that all studies are based on the
OGLE data.

The numbers from Table~4 are visualized in Fig.~14 -- for the LMC in the
left panel and for the SMC in the right panel, both on the ZW84 scale. All
SMC [Fe/H] values from Table~4 have been transformed from the J95 to the
ZW84 metallicity scale with the same equation (by J95), including the value
of Haschke \etal (2012), which was originally calculated using the equation
of Papadakis \etal (2000), and that original value is provided in
Table~4. We do not recalculate ${\rm [Fe/H]}_{ZW84}$ for the LMC, because
the difference between the transformations of J95 and Papadakis \etal
(2000) is very small.

In the case of the LMC all [Fe/H] estimates from different methods are well
consistent and the mean value is ${\rm [Fe/H]}_{ZW84}=-1.59$~dex, which
corresponds to ${\rm [Fe/H]}_{J95}=-1.39$~dex. This is because the
metallicity of the LMC is in the ``safe'' range, where all methods give
roughly similar results. In the case of the more metal-poor SMC, there is a
noticeable discrepancy between values from the literature and the result of
this study, which is expected as a consequence of Fig.~7: median
metallicity from Haschke \etal (2012) with S05 method is the highest, while
the result of this study with N13 method is the lowest.

\subsection{Metallicity Gradient}
The metallicity gradient in the LMC has been found in asymptotic giant
branch (AGB) stars (derived by Cioni 2009) and from OGLE-III RRab stars
(Feast \etal 2010, Wagner-Kaiser and Sarajedini 2013). No metallicity
gradient was found in the red giant branch star population in 28 populous
LMC clusters (Grocholski \etal 2006) and in the outer disk of the LMC
(Carrera \etal 2011). Piatti and Geisler (2013) also report no metallicity
gradient in their analysis of an age--metallicity relationship of the LMC
field star population. Interestingly, Haschke \etal (2012) and Deb and
Singh (2014) did not detect a metallicity gradient in the OGLE-III RRab
data.

Several authors, who report no gradient, argue that the errors on the mean
metallicity values in distance bins are much larger than the gradient
itself, thus questioning its existence. However, our study contains a much
more extended area around the LMC, out to $\approx15$~kpc, and we observe a
consistent decrease in median [Fe/H] over that range. In addition, there is
a clear evidence that the proportion of metal-rich to metal-poor RRab stars
decreases with distance from the center, giving another argument for the
metallicity gradient.

Our results for the inner and outer parts of the LMC are consistent with
those of Wagner-Kaiser and Sarajedini (2013), who found a gradient with a
slope of $-0.027\pm0.002$~dex/kpc within the inner 5~kpc. On the other had,
the slope we find is steeper than the slope of $-0.015\pm0.003$~dex/kpc
from Feast \etal (2010), when taking into account the inner parts of the
galaxy, but consistent within errors, if we take the total gradient value
for the LMC of $-0.019\pm0.002$~dex/kpc. Both studies were based on OGLE
RR~Lyr stars. Cioni (2009) found a decrease of iron abundance of
$-0.047\pm0.003$~dex/kpc out to $\approx8$~kpc, derived from the
asymptotic giant branch (AGB) stars, which is steeper than found in this
study. Interestingly, they noticed that the AGB gradient at $5<r<8$~kpc is
marginally flatter than in the inner disk, which is consistent with the
transition between inner disk and outer disk/halo components.  We see this
break with RRab stars, whose [Fe/H] gradient actually flattens in that
distance range.

In the case of the SMC, the metallicity gradient was found among 350 red
giant branch stars in 13 fields in different positions out to
$\approx4\arcd$ from the SMC center (Carrera \etal 2008) and among 3037 red
giants spread across approximately 37.5~deg$^2$ centered on this galaxy
(Dobbie \etal 2014). Kapakos and Hatzidimitriou (2012) also found a
metallicity gradient among OGLE-III RRab stars. Contrary, Cioni (2009) --
from AGB stars, Parisi \etal (2009) -- from the study of 16 SMC clusters,
and Deb \etal (2015) -- from OGLE-III RRab sample, report no gradient, and
we support this result with our findings.

\Section{Summary}
We used the OGLE-IV collection of fundamental mode RR~Lyr stars to find a
new relation between the phase parameter $\varphi_{31}$ in the {\it V}- and
{\it I}-band, in order to facilitate [Fe/H] calculation for large datasets
of {\it I}-band RRab light curves. The relation is non-linear and depends
on metallicity, so its applicability is limited. If the
metallicity-independent version of the relation is used, the final [Fe/H]
distribution is subject to systematic errors at the high and low ends, such
that the low metallicity values are overestimated and high metallicity
values are underestimated. The median of the [Fe/H] distribution is not
affected.

The comparison of the new photometric metallicity calculation method by
Nemec \etal (2013) with the widely used methods of Jurcsik and Kovacs
(1996) and Smolec (2005) shows that the problem of overestimating [Fe/H] at
low metallicity values by the two latter methods is non negligible, when
the metallicity-dependent ($\varphi^I_{31}\rightarrow\varphi^V_{31}$)
transformation is used together with the Nemec \etal (2013)
method. However, when the metallicity-independent ($\varphi^I_{31}
\rightarrow\varphi^V_{31}$) transformation is applied, the differences are
smaller.

We use the method of Nemec \etal (2013) to construct a photometric
metallicity map of the Magellanic System from OGLE-IV RRab using the
metallicity-independent ($\varphi^I_{31}\rightarrow\varphi^V_{31}$)
relation. We find that the mean iron abundance in the LMC is
$-1.39\pm0.44$~dex, which is consistent with previous findings. In the case
of the SMC, mean [Fe/H] is $-1.77\pm0.48$~dex and it is lower than values
from the literature, which is explained by larger discrepancies between
methods of [Fe/H] calculation at low metallicities, such as those observed
in the SMC.

While there appears to be no, or very mild, metallicity gradient in the
SMC, we clearly see that in the LMC [Fe/H] decreases with distance from the
galaxy center. The decline is evident within 4~kpc from the center, then
the metallicity remains constant until about 8~kpc, when it shows another
slight decline out to $\approx13$~kpc. The slope of the gradient is
$-0.029\pm0.002$~dex/kpc in the inner 5~kpc and $-0.030\pm0.003 $~dex/kpc
beyond 8~kpc. The total slope, for the entire LMC, is
$-0.019\pm0.002$~dex/kpc.

\Acknow{We would like to thank the anonymous Referee for remarks that
greatly improved this paper. D.M.S. is supported by the Polish National
Science Center (NCN) under the grant no. 2013/11/D/ST9/03445 and the
Polish Ministry of Science and Higher Education under the grant
``Iuventus Plus'' No. 0420/IP3/2015/73.  The OGLE project has received
funding from the NCN grant MAESTRO 2014/14/A/ST9/00121 to AU.}


\begin{references}
\refitem{Alcock, C.J., \etal}{2000}{\AJ}{119}{2194}
\refitem{Benk\H{o}, J.M., \etal}{2010}{\MNRAS}{409}{1585}
\refitem{Braga, V.F., \etal}{2015}{\ApJ}{799}{165}
\refitem{Carrera, R., Gallart, C., Aparicio, A., Costa, E., M\'endez, R.A., and No\"el, N.E.D.}{2008}{\AJ}{136}{1039}
\refitem{Carrera, R., Gallart, C., Aparicio, A., and Hardy, E.}{2011}{\AJ}{142}{61}
\refitem{Carretta, E., Bragaglia, A., Gratton, R., D'Orazi, V., and Lucatello, S.}{2009}{\AA}{508}{695}
\refitem{Cioni, M.-R.L.}{2009}{\AA}{506}{1137}
\refitem{Deb, S., and Singh, H.P.}{2010}{\MNRAS}{402}{691}
\refitem{Deb, S., and Singh, H.P.}{2014}{\MNRAS}{438}{2440}
\refitem{Deb, S., Singh, H.P., Kumar, S., and Shashi M.}{2015}{\MNRAS}{449}{2768}
\refitem{Dobbie, P.D., Cole, A.A., Subramaniam, A., and Keller, S.}{2014}{\MNRAS}{442}{1680}
\refitem{Feast, M.W., Abedigamba, O.P., and Whitelock, P.A.}{2010}{\MNRAS}{408}{76}
\refitem{Gonzalez, O.A., Rejkuba, M., Zoccali, M., Valent, E., Minniti, D., and Tobar, R.}{2013}{\AA}{552}{110}
\refitem{Grocholski, A.J., Cole, A.A., Sarajedini, A., Geisler, D., and Smith, V.V.}{2006}{\AJ}{132}{1630}
\refitem{Haschke, R., Grebel, E.K., Duffau, S., and Jin, S.}{2012}{\AJ}{143}{48}
\refitem{Jeon, Y.-B., Ngeow, C.-C., and Nemec, J.M.}{2014}{IAU Symp.}{301}{427}
\refitem{Jurcsik, J.}{1995}{\Acta}{45}{653}
\refitem{Jurcsik, J., and Kov\'acs, G.}{1996}{\AA}{312}{111}
\refitem{Kapakos, E., Hatzidimitriou, D., and Soszyñski, I.}{2011}{\MNRAS}{415}{1366}
\refitem{Kapakos, E., and Hatzidimitriou, D.}{2012}{\MNRAS}{426}{2063}
\refitem{Kinemuchi, K., \etal}{2006}{\AJ}{132}{1202}
\refitem{Kov\'acs, G., and Zsoldos, E.}{1995}{\AA}{293}{L57}
\refitem{Nemec, J.M., \etal}{2013}{\ApJ}{773}{181}
\refitem{Papadakis, I., Hatzidimitriou, D., Croke, B.F.W., and Papamastorakis, I.}{2000}{\AJ}{119}{851}
\refitem{Parisi, M.C., Grocholski, A.J., Geisler, D., Sarajedini, A., and Clari\'a, J.J.}{2009}{\AJ}{138}{517}
\refitem{Piatti, A.E., and Geisler, D.}{2013}{\AJ}{145}{17}
\refitem{Pietrukowicz, P., \etal}{2012}{\ApJ}{750}{169}
\refitem{Pietrukowicz, P., \etal}{2015}{\ApJ}{811}{113}
\refitem{Sandage, A.}{2004}{\AJ}{128}{858}
\refitem{Sans Fuentes, S.A., and De Ridder, J.}{2014}{\AA}{571}{59}
\refitem{Smolec, R.}{2005}{\Acta}{55}{59}
\refitem{Soszyñski, I., \etal}{2009}{\Acta}{59}{1}
\refitem{Soszyñski, I., \etal}{2011}{\Acta}{61}{1}
\refitem{Soszyñski, I., \etal}{2014}{\Acta}{64}{177}
\refitem{Soszyñski, I., \etal}{2016}{\Acta}{66}{131}
\refitem{Szczygie³, D., Pojmañski, G., and Pilecki, B.}{2009}{\Acta}{59}{137}
\refitem{Torrealba, G., \etal}{2015}{\MNRAS}{446}{2251}
\refitem{Udalski, A., Szymañski, M.K., and Szymañski, G.}{2015}{\Acta}{65}{1}
\refitem{van der Marel, R.P., and Cioni, M.-R.L.}{2001}{\AJ}{122}{1807}
\refitem{Wagner-Kaiser, R., and Sarajedini, A.}{2013}{\MNRAS}{431}{1565}
\refitem{Weinberg, M.D., and Nikolaev, S.}{2001}{\ApJ}{548}{712}
\refitem{Zinn, R., and West, M.J.}{1984}{\ApJS}{55}{45}
\end{references}
\end{document}